\documentclass[aps,prl,reprint,superscriptaddress,floatfix,nofootinbib]{revtex4-1}
\usepackage[dvips]{graphicx}
\usepackage[utf8]{inputenc}
\usepackage{subfigure}
\usepackage{amssymb}
\usepackage{array,multirow}
\usepackage{amsmath}
\usepackage{mathrsfs}
\usepackage{fancyhdr}
\usepackage{verbatim}
\usepackage{color}
\usepackage{enumitem}
\usepackage{appendix}
\usepackage{amssymb}
\usepackage{amsthm}
\usepackage{amsmath}

\newcommand{\lat}{\mathrm{lat}}

\newcommand{\cont}{\mathrm{cont}}

\newcommand{\MSb}{\overline{\mathrm{MS}}}

\newcommand{\chidof}{\chi^2/{\rm d.o.f.}}

\begin{document}

\title{Running coupling constant from position-space current-current correlation functions in three-flavor lattice QCD}

\author{Salvatore Cal\`i}
\email[]{salvatore.cali@uj.edu.pl}
\affiliation{Institute of Theoretical Physics, Jagiellonian University, ul. \L ojasiewicza 11, 30-348 Krak\'ow, Poland}
\author{Krzysztof Cichy}
\email[]{krzysztof.cichy@gmail.com}
\affiliation{Faculty of Physics, Adam Mickiewicz University, ul.\ Uniwersytetu Pozna\'nskiego 2, 61-614 Pozna\'n, Poland}
\author{Piotr Korcyl}
\email[]{piotr.korcyl@uj.edu.pl}
\affiliation{Institute of Theoretical Physics, Jagiellonian University, ul. \L ojasiewicza 11, 30-348 Krak\'ow, Poland}
\affiliation{Institut f\"ur Theoretische Physik, Universit\"at Regensburg, D-93040 Regensburg, Germany}
\author{Jakob Simeth}
\email[]{jakob.simeth@ur.de}
\affiliation{Institut f\"ur Theoretische Physik, Universit\"at Regensburg, D-93040 Regensburg, Germany}

\date{\today}

\begin{abstract}
In this Letter, we provide a determination of the coupling constant in three-flavor quantum chromodynamics (QCD), $\alpha^{\MSb}_s(\mu)$, for $\MSb$ renormalization scales $\mu \in (1,\,2)$ GeV.
The computation uses gauge field configuration ensembles with $\mathcal{O}(a)$-improved Wilson-clover fermions generated by the Coordinated Lattice Simulations (CLS) consortium. 
Our approach is based on current-current correlation functions and has never been applied before in this context.
We convert the results perturbatively to the QCD $\Lambda$-parameter and obtain $\Lambda_{\MSb}^{N_f=3} = 342 \pm 17$ MeV, which agrees with the world average published by the Particle Data Group and has competing precision. 
The latter was made possible by a unique combination of state-of-the-art CLS ensembles with very fine lattice spacings, further reduction of discretization effects from a dedicated numerical stochastic perturbation theory simulation, combining data from vector and axial-vector channels and matching to high-order perturbation theory.
\end{abstract}


\maketitle


\noindent\textit{Motivation}: 
The strength of strong interactions, parametrized by the scale-dependent coupling $\alpha_s$, typically quoted at the $Z$-boson pole mass, is one of the most important parameters of the Standard Model (SM). 
It is required in perturbative calculations in collider physics and its uncertainty is one of dominant sources of uncertainty in several SM predictions, as well as in tests of SM extensions \cite{Salam:2017qdl}. 
Due to the non-Abelian Yang-Mills nature of quantum chromodynamics (QCD), $\alpha_s$ vanishes asymptotically at very high energies \cite{Gross:1973id,Politzer:1973fx} and experiments are able to follow this energy dependence in various processes over a wide range of energy scales.
This allows to determine $\alpha_s$ at several scales by fitting experimental data and matching to a perturbative expansion of an appropriate observable.
Equivalently, using renormalization group concepts, one may parametrize the running of $\alpha_s$ by a single parameter, $\Lambda$, corresponding to the scale where perturbation theory breaks down.
Examples of experimental processes for the extraction of the strong coupling or the $\Lambda$-parameter are hadronic $\tau$ decays, deep inelastic scattering and hadronic final states of $e^+e^-$ annihilation. For a review of many aspects of such determinations and the obtained values, see the Particle Data Group (PDG) review \cite{Tanabashi:2018oca}.
However, the strong coupling constant or the $\Lambda$-parameter can also be extracted directly from the QCD Lagrangian, using the non-perturbative formulation of QCD on the lattice.
This proceeds by calculating appropriately designed short-distance Euclidean observables and, again, matching them to their perturbative expansions.
Over the years, several methods how to design such observables have been proposed.
Recent investigations employed e.g.\ step scaling methods \cite{Bruno:2017gxd,Ishikawa:2017xam}, the static quark-antiquark potential \cite{Husung:2017qjz,Karbstein:2018mzo,Takaura:2018lpw}, the vacuum polarization function \cite{Hudspith:2018bpz}, the heavy-quark current two-point correlation function \cite{Maezawa:2016vgv}, QCD vertices (e.g.\ ghost-gluon) \cite{Zafeiropoulos:2019flq} or eigenvalues of the lattice Dirac operator \cite{Nakayama:2018ubk}.
For a discussion and overview of these and older results, see the Flavor Lattice Averaging Group (FLAG) review \cite{Aoki:2019cca}.
The determinations from experiments and from the lattice enter the world average of $\alpha_s$ in the PDG review \cite{Tanabashi:2018oca}, recently with a visibly larger impact of lattice results due to their smaller total uncertainties.

In this Letter, we describe a novel method of estimating the running of the coupling or the $\Lambda$-parameter, using numerical simulations of QCD.
The proposed method employs large volume simulations, it has a moderate numerical cost and is clean and straightforward from the theoretical point of view. It is based on current-current correlation functions in position space, objects well studied and easily accessible in the lattice QCD framework. Thanks to the combination of very fine lattices generated by the Coordinated Lattice Simulations (CLS) effort \cite{Bruno:2014jqa,Bali:2016umi}, precise renormalization factors from the chirally rotated Schr\"odinger functional ($\chi$SF) framework \cite{DallaBrida:2018tpn} and $\mathcal{O}(a)$-improvement coefficients \cite{Bulava:2015bxa,Heitger:2017njs}, subtraction of leading-order and next-to-leading-order discretization effects estimated in the numerical stochastic perturbation theory (NSPT) formulation \cite{DiRenzo:2004hhl} and various improved analysis techniques, it yields a competitive total uncertainty. 
Therefore, it may serve as a robust method of estimating the $\Lambda$-parameter. 
The approach presented here is not limited to $\Lambda$ -- a position-space analysis may also be used to reliably estimate other important observables, such as quark and gluon condensates \cite{Tomii:2017cbt}, quark masses \cite{Tomii:2018zix} or operator renormalization functions \cite{Gimenez:2004me,Cichy:2012is,Cichy:2016qpu,Tomii:2016xiv}.

\noindent\textit{Strategy}:
The strategy proposed in this Letter uses a combination of numerical lattice QCD calculations and high-order perturbative results. We concentrate on correlation functions of flavor non-singlet bilinear quark operators of the form
\begin{equation}
C_{\Gamma}(1/x,m_q,a) = Z_\Gamma^2 \langle \bar{\psi}^i(x) \Gamma \psi^j(x) \bar{\psi}^j(0) \Gamma \psi^i(0) \rangle, 
\label{eq. correlator}
\end{equation}
where $x$ is the physical distance, $m_q$ is the quark mass of degenerate three flavors of quarks, $a$ denotes the lattice spacing, $\Gamma = \{ V \equiv \gamma_{\mu}, A \equiv \gamma_{\mu} \gamma_5 \}$, $i \ne j$ and $Z_\Gamma$ is the (scale-independent) renormalization factor. 
For a reliable extraction of $\alpha_s$, we need to work in the regime of distances satisfying a window condition, $a\ll x \ll \Lambda^{-1}$.
The former condition guarantees that discretization effects are not enhanced, while the latter establishes that reliable contact to perturbation theory can be made.
After extrapolating the correlation functions to the continuum limit and after renormalization, they can be matched to their perturbative expansions in terms of $\alpha_s$, typically in the $\MSb$ scheme ($\alpha_{\MSb}(\mu)$),
\begin{equation}
    C_{\Gamma}(\mu) = c^{(1)}_{\Gamma} \alpha_{\MSb}(\mu) + c^{(2)}_{\Gamma} \alpha_{\MSb}^2(\mu) + \dots,
    \label{eq. strategy}
\end{equation}
where $C_{\Gamma}(\mu)\equiv C_{\Gamma}(1/x,m_q=0,a=0)$.
Such an expansion of current-current correlators is presently available up to 4 loops \cite{Chetyrkin:2010dx}.
Knowing $C_{\Gamma}(\mu)$ from numerical simulations and the analytic form of coefficients $c^{(i)}_{\Gamma}$, we solve Eq.~\eqref{eq. strategy} for $\alpha_{\MSb}(\mu)$. Subsequently, we convert that value to our estimate of the $\Lambda$-parameter. 
Now, we provide details of the different steps needed to reliably obtain $C_{\Gamma}(\mu)$.

\noindent\textit{Crucial elements of the analysis}: 
We start with the bare lattice data for correlation functions $C_{A/V}(1/x,m_q,a)$ with the $\mathcal{O}(a)$-improvement of the currents implemented by using improvement coefficients $c_A$ from Ref.~\cite{Bulava:2015bxa} and $c_V$ from Ref.~\cite{Heitger:2017njs}. The ensembles used in this study are summarized in Tab.~\ref{tab. ensembles}.
We perform 64 inexact and 2 exact measurements per configuration using the truncated solver method \cite{Bali:2009hu} and for every lattice distance $x/a$, we average correlators evaluated from all sites equivalent with respect to the hypercubic symmetry of the lattice.
\begin{table}
\begin{center}
\begin{ruledtabular} 
\begin{tabular}{cccccc}
$\beta$ & name & $\kappa_l = \kappa_s$ & $m_{\pi}$ [MeV] & $t_0/a^2$ & \# conf.\\
\hline
3.46 & B450   & 0.136890 & 419 & 3.663(11)  & 320 \\
3.46 & rqcd30 & 0.136959 & 320 & 3.913(15) & 280 \\
3.46 & X450 & 0.136994 & 264 & 3.994(10) & 280 \\
3.55 & B250   & 0.136700 & 709 & 4.312(8)   & 84  \\
3.55 & N202   & 0.137000 & 412 & 5.165(14)  & 177 \\
3.55 & X250   & 0.137050 & 348 & 5.283(27) & 182 \\
3.55 & X251   & 0.137100 & 269 & 5.483(26) & 177 \\
3.7  & N303   & 0.136800 & 641 & 7.743(23) & 99  \\
3.7  & N300   & 0.137000 & 423 & 8.576(21) & 197 \\
3.85 & N500   & 0.13672514 & 599 &  12.912(75) & 100 \\
3.85 & J500   & 0.136852 & 410 & 14.045(38) & 120 \\
\end{tabular}
\end{ruledtabular} 
\end{center}
\caption{Subset of $N_f=2+1$ CLS ensembles along the symmetric line $\kappa_l=\kappa_s$ used in this work \cite{Bruno:2014jqa,Bali:2016umi}. The gauge action is tree-level Symanzik-improved, while the fermionic one is the Wilson $O(a)$-improved (clover) action with the improvement coefficient, $c_{SW}$, determined non-perturbatively. 
rqcd30, X450, B450, X250, X251 have been generated by the RQCD and Mainz collaborations. 
For more details, see 
Refs.~\cite{Bruno:2014jqa,Bali:2016umi}. The values of $t_0/a^2$ are the reweighted estimates using the symmetric definition of the Yang-Mills action density \cite{inprep2}.
The lattice spacings corresponding to different $\beta$ values are 0.07582(24) fm ($\beta=3.46$), 0.0644(7) fm ($\beta=3.55$), 0.0499(5) fm ($\beta=3.7$) and 0.0391(15) fm ($\beta=3.85$) \cite{inprep2}. 
The scale setting is based on the determination of light hadron masses on all CLS ensembles with 6 lattice spacings from $\approx0.1$ fm down to below 0.04 fm and pion masses $\in[135,420]$ MeV. The Wilson flow scales are determined in the continuum using the $\Upsilon$ baryon mass as input.
The last column indicates the number of configurations used. \label{tab. ensembles}}
\end{table}

At fixed lattice spacing and lattice distance, we extrapolate the correlators to the chiral limit.
We use a fitting ansatz linear in the dimensionless combination $y = t_0 m^2_{\pi}$, where $t_0$ is an intermediate unphysical scale introduced in Ref.~\cite{Luscher:2010iy} and we take the values of $t_0/a^2$ from Ref.~\cite{inprep2}. 
The quality of the chiral fit was tested at $\beta=3.55$, where we have four pion masses available.
We compared the linear fit in $y$ to either all (``lin4'') or the 3 lightest masses (``lin3'') with the quadratic one to all masses (``quad4'') for all the relevant distances (for more details, see the supplement). The small differences that we observe in the chiral limit amount, on average, to 0.17\% (lin3 vs.\ lin4) and 0.26\% (quad4 vs.\ lin4) at the level of correlators. Conservatively, we propagate the latter to $\alpha_s$ via a bootstrap procedure, taking for other $\beta$-values the linear ansatz.
We denote the massless correlator by $C_{\Gamma}(1/x,a)$.
The massless correlators are then expressed in the $\MSb$ scheme, using renormalization factors calculated in Ref.~\cite{DallaBrida:2018tpn}, determined using the $\chi$SF framework \cite{Sint:2010eh}.

\begin{figure}
\includegraphics[width=0.465\textwidth]{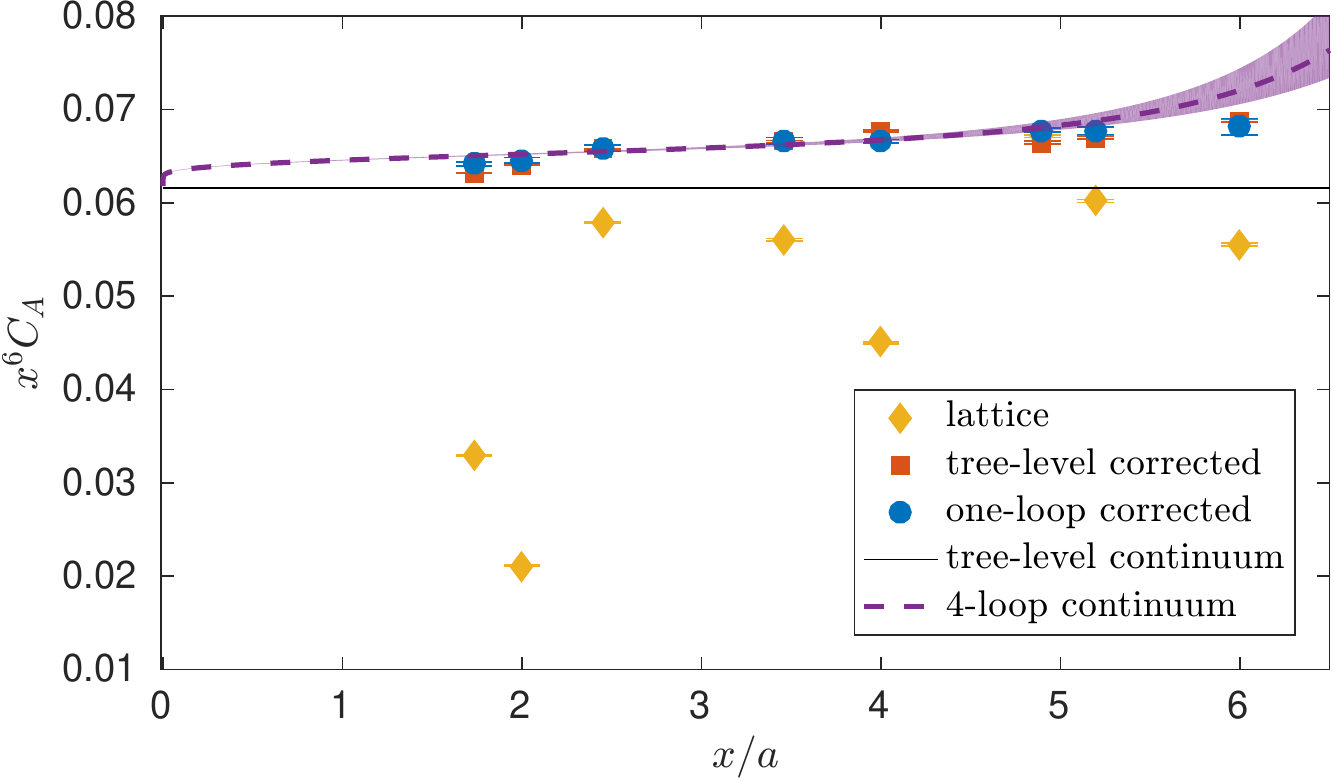}
\caption{Impact of the tree-level (red squares) and 1-loop (blue circles) improvement of the massless axial current-current correlation function at $\beta=3.85$. The unimproved lattice correlators are shown as yellow rhombi. Shown are also tree-level and 4-loop \cite{Chetyrkin:2010dx} continuum perturbative lines. The difference of the latter, corresponding to $\Lambda_{\MSb}^{N_f=3}$ found in this work, with respect to blue points isolates the remaining discretization effects at this $\beta$.
\label{fig: one-loop}}
\end{figure}

A significant step to reliably perform the continuum limit extrapolation is to reduce the size of discretization effects present in the data.
To this aim, we perturbatively compute $O(a^\infty g^2)$ artifacts,
i.e.\ we replace the correlation functions
\begin{equation}
C_{\Gamma}(1/x,a) \to C_{\Gamma}(1/x,a) + \big( C^{\textrm{cont}}_{\Gamma}(1/x) - C^{\textrm{lat}}_{\Gamma}(a/x) \big),
\label{eq: improvement}
\end{equation}
where $C^{\textrm{t}}_{\Gamma} = C^{\textrm{free,\,t}}_{\Gamma} + g_0^2 C^{\textrm{1-loop,\,t}}_{\Gamma}$,  $\textrm{t}\in \{ \textrm{cont}, \textrm{lat} \}$ and the superscript free/1-loop denotes the tree-level/1-loop contributions.
The massless $\MSb$ continuum correlators, $C^{\textrm{cont}}_{A/V}$, are given in Ref.~\cite{Chetyrkin:2010dx}.
In turn, $C^{\textrm{lat}}_{A/V}$ are computed in NSPT \cite{DiRenzo:2004hhl} along the lines of Refs.~\cite{Simeth:2013ima,Simeth:2015xua}\footnote{A more detailed description of the NSPT calculation will be presented in a separate publication~\cite{jakob}. For a shorter account, see the supplement.}
and are expressed in the $\MSb$  scheme using the renormalization factors for the employed gauge action \cite{Taniguchi:1998pf}.
Thus, all terms appearing on the RHS of Eq.~\eqref{eq: improvement} are correctly normalized correlators in the same scheme. 
The improved correlator, hence, has leading cutoff effects of $\mathcal{O}(a^2 g^4)$. 
We demonstrate the reduction of discretization effects in Fig.~\ref{fig: one-loop}, depicting the distance dependence of $C_A(1/x,a)$ at $\beta=3.85$, for all points used in the extraction of $\alpha_s$.
We show three data sets: without any correction, with the tree-level correction only and with the full 1-loop NSPT correction. 
The scatter of data points is clearly reduced, yielding a smooth curve.
It is important to emphasize that the tree-level corrected data, even though seemingly already smooth, still prohibit any meaningful extraction of $\alpha_s$ (see supplement for more details).
Thus, reliable control of discretization effects necessitates the use of the 1-loop subtraction of artifacts (all orders in the lattice spacing) and this step is crucial for the success of the method.
Note also that the 1-loop correction is drastically smaller than the tree-level one, hinting at good convergence of this expansion.
Moreover, the 1-loop-corrected correlators are very close to the 4-loop continuum perturbative curve \cite{Chetyrkin:2010dx}, indicating that the remaining discretization effects are small at this lattice spacing.
  
In order to perform the continuum extrapolations, we need to follow the lines of constant physics. In our case, the only relevant scale is the correlator distance $x$, which we keep fixed in physical units by interpolating to the desired distance at all $\beta$ values.
We use two interpolation ansatzes, linear and quadratic in $x^2$, between the two and three closest data points to find the interpolated value at each lattice spacing. We consider three lattice (``democratic'') directions, for which hypercubic artifacts are known to be the smallest \cite{Cichy:2012is,Cichy:2016qpu}:
$(0,k,k,k),\- (k,k,k,k),\- (0,k,k,2k)$ with $k \in \{1,2,3\}$ and interpolate independently for each of them.
We note that other types of points do not make it possible to extract $\alpha_s$ at sufficiently small distances or break the rotational symmetry too severely (``non-democratic'' directions).
Hence, similarly as in momentum-space studies of renormalization functions (see e.g.\ Ref.~\cite{Alexandrou:2015sea}), the 1-loop subtraction needs to be supplemented by a ``democratic'' criterion (see also the supplement).
In this way, we keep the discretization effects related to the breaking of rotational symmetry well-controlled and fixed as we change the lattice spacing. We use the difference of the two interpolation models as the systematic uncertainty associated to this step. 
    
If discretization effects are under control, the continuum limits corresponding to the same physical distance should agree for each of the three lattice directions. 
We checked that this is the case and hence, we performed combined continuum fits of data for all three directions.
Depending on the distance (and, thus, the available lattice spacings), we use from 6 to 12 data points and constrain the fit by a common value in the continuum, $C_\Gamma(1/x,a=0)$. 
The fitting ansatz reads
\begin{equation}
C_\Gamma(1/x,a) = C_\Gamma(1/x,a=0) + \sum_{i = \textrm{lattice direction}} \alpha_i a^2 
\end{equation}
and has 4 fit parameters.

\begin{figure}
\includegraphics[width=0.465\textwidth]{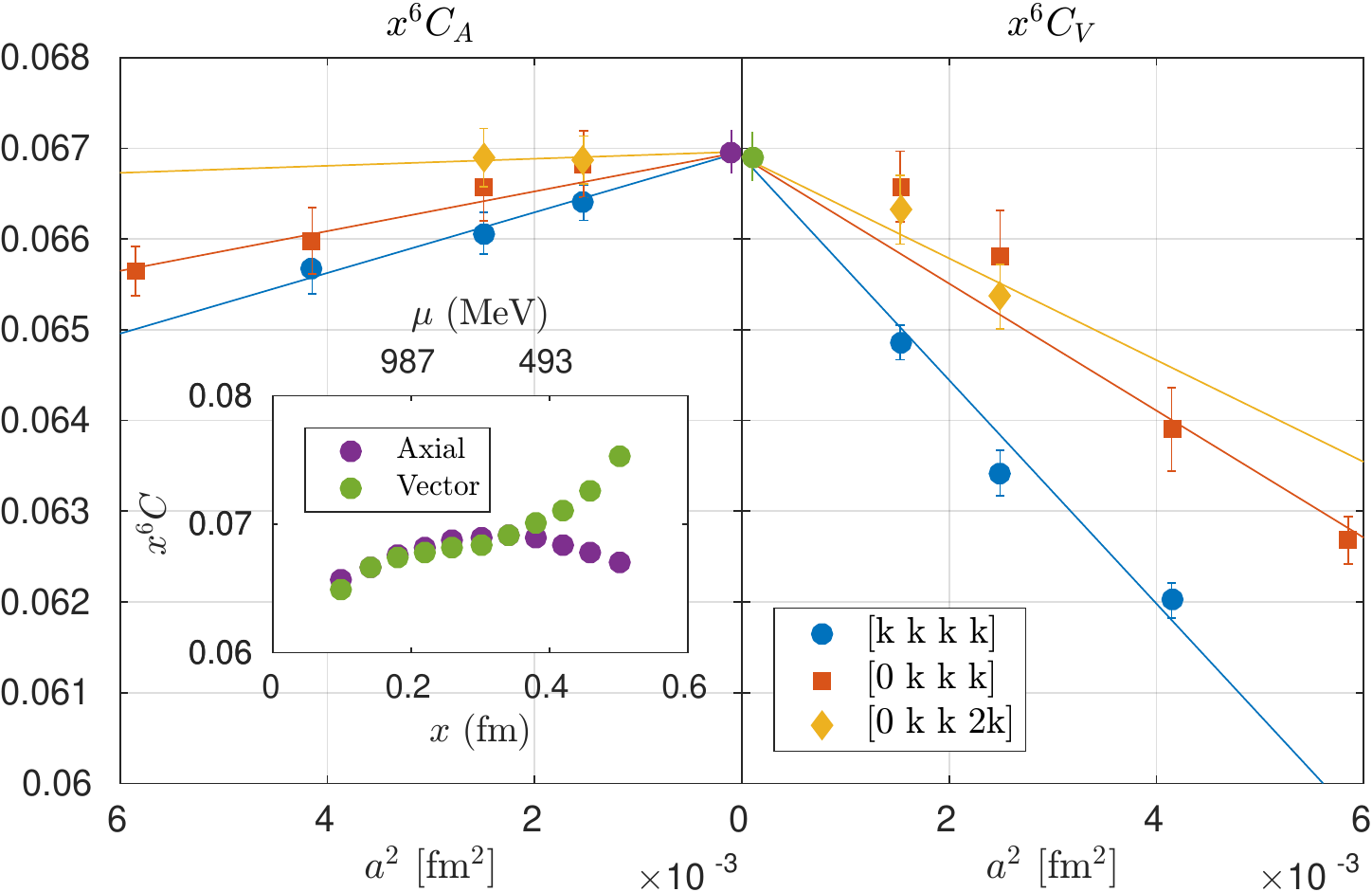}
\caption{Continuum extrapolation of the axial (left) and vector (right) correlators at $x=0.15$ fm. Both extrapolations are performed independently, but their extrapolated values agree within uncertainties. The inset shows the distance dependence of the continuum-extrapolated axial and vector correlators. The error bars are smaller than the symbol size. Within our uncertainties, the correlators exhibit agreement for distances $\lesssim0.35$ fm.
\label{fig: continuum}}
\end{figure}

The difference between the axial and vector correlation functions was estimated in various frameworks, for a review see Ref.~\cite{RevModPhys.65.1}, including lattice QCD \cite{PhysRevD.64.094508,Tomii:2017cbt}. Also empirical data exist for this observable \cite{aleph}. At short distances, the difference between the vector and axial correlators is reliably provided by the operator product expansion \cite{SHIFMAN1979385}. Using estimates from Ref.~\cite{PhysRevLett.86.3973}, the relative difference ranges from 0.03\% at $x=0.1$ fm up to 1.5\% at $x=0.3$ fm. Hence, within the statistical and systematic precision of our data, the two correlators are indistinguishable in that range of distances, see the inset of Fig.~\ref{fig: continuum} and the supplement for more details.
We use this observation in a two-fold way. 
a) First, we employ it as a test of the reliability of the continuum extrapolation. 
In further analysis, we consider only the physical distances for which the independently extrapolated axial and vector correlators agree within their uncertainties. 
On the one hand, this criterion excludes lattice directions and physical distances which are too short and no control over discretization effects is possible, setting the lower limit to 0.1 fm.
On the other hand, the two correlators are no longer equivalent at distances larger than around 0.35 fm within our precision, which sets the upper limit on the physical distances where the impact of non-perturbative condensates is negligible.
Note that the scale where the correlators become incompatible is related to effects of spontaneous chiral symmetry breaking and not to the breakdown of perturbation theory.
In Fig.~\ref{fig: continuum}, we show an example of the continuum extrapolations of the axial and vector correlators at the physical distance of $x=0.15$ fm. 
For examples for other distances, see the supplement.
The fits are performed independently for both Dirac structures and in both cases, the combined fits to our three lattice directions provide a good description of the data, which holds also at other relevant distances (with $\chidof\in[0.1,2]$). We emphasize that this is achieved only in the case of NSPT-corrected data, while continuum fits for only tree-level corrected data lead to $\chidof\approx10-20$ at the relevant distances.
Moreover, although the individual data points at finite lattice spacing are different for different Dirac structures, $C_A=C_V$ in the continuum. b) Second, for the physical distances in the relevant range 0.1-0.2 fm, we use the independent data for $C_A$ and $C_V$ and consider their average, thus gaining in statistical precision.

\begin{figure}
\includegraphics[width=0.465\textwidth]{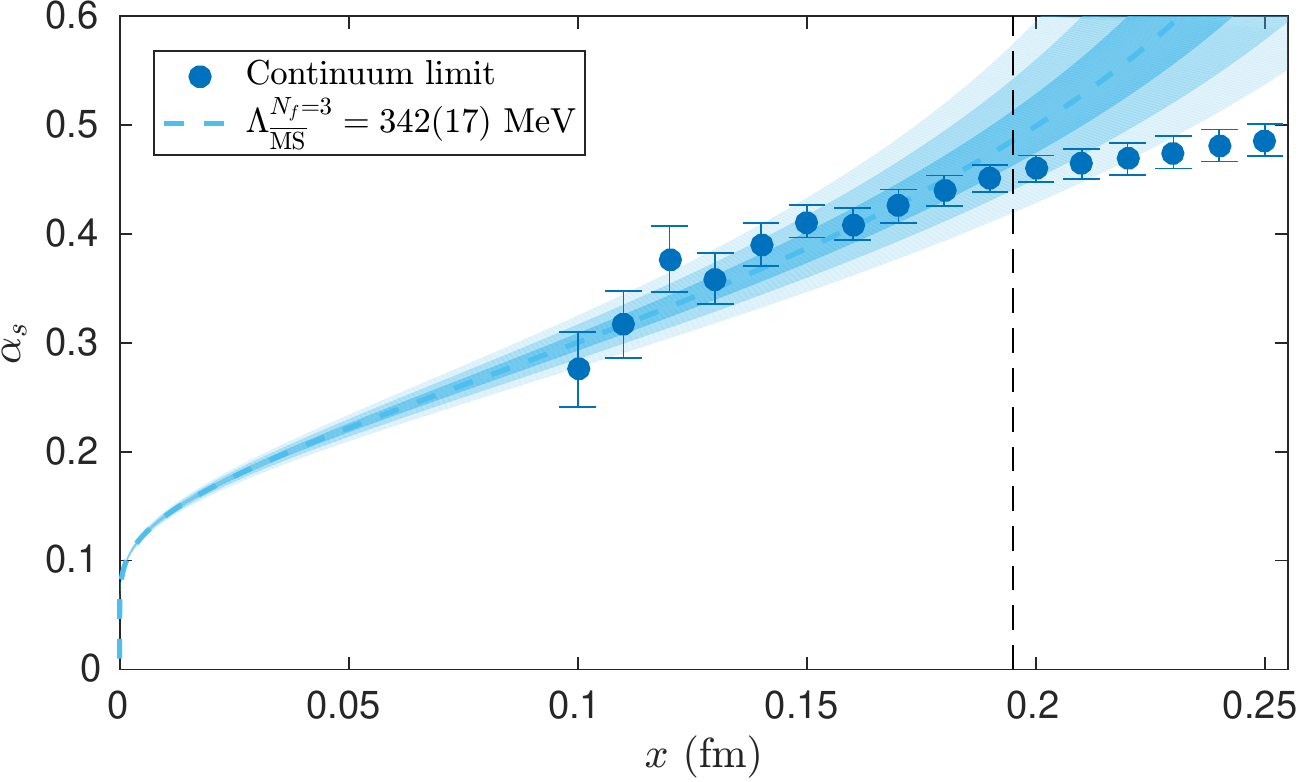}
\caption{Running of $\alpha_s$ extracted from the lattice data (blue points). At distances above around 0.2 fm, the running freezes and indicates the breakdown of matching to perturbation theory. The shaded blue band is the corresponding 5-loop perturbative running \cite{Baikov:2016tgj,Luthe:2016ima,Herzog:2017ohr}  using the final value of $\Lambda^{N_f=3}_{\MSb}$ determined in this work (see section \emph{Final result}). The darkest band corresponds to the 1-$\sigma$ total uncertainty of $\Lambda^{N_f=3}_{\MSb}$ and the lighter bands to 2-$\sigma$ and 3-$\sigma$.
\label{fig: alpha results}}
\end{figure}

\begin{figure}
\includegraphics[width=0.465\textwidth]{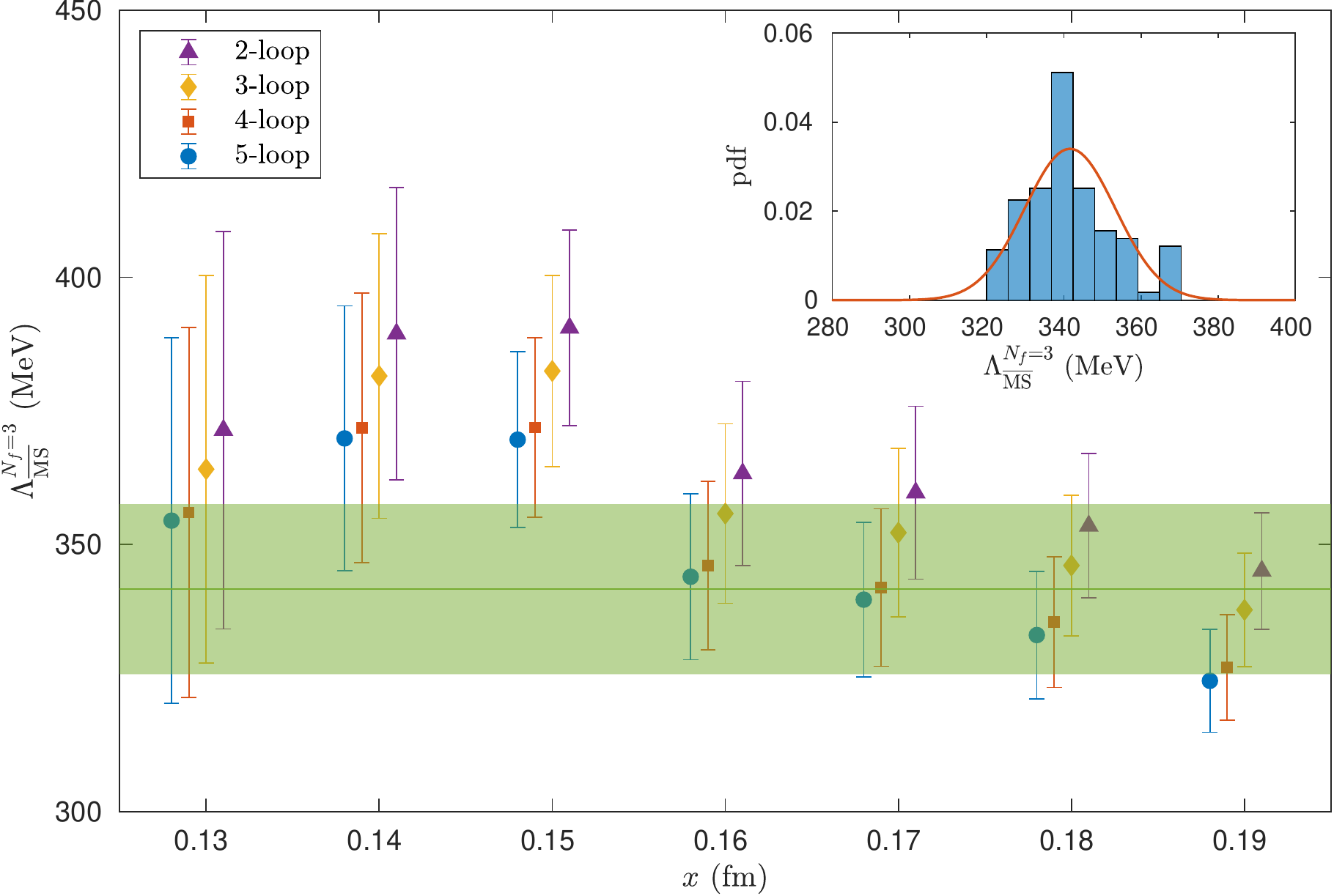}
\caption{Results for $\Lambda_{\MSb}^{N_f=3}$ obtained from $\alpha_s$ at different distances. Shown are conversions between $\alpha_s$ and $\Lambda$ \cite{Callan:1970yg,Symanzik:1970rt} using from 2- to 5-loop perturbative $\beta$-functions \cite{Baikov:2016tgj,Luthe:2016ima,Herzog:2017ohr}. Our final value (indicated by the green band whose width is the total uncertainty) is obtained by a systematic procedure explained in the text.
The inset shows the histogram of results corresponding to different ranges of distances taken into account in the systematic procedure, along with a Gaussian fit.
\label{fig: final}}
\end{figure}

Having the continuum-extrapolated $\MSb$ correlators, we know both sides of Eq.~\eqref{eq. strategy} and we can determine $\alpha_{\MSb}(\mu)$ for different scales, corresponding to different physical distances $1/\mu=x$. 
The results are shown in Fig.~\ref{fig: alpha results}. At distances above around 0.2 fm (scales below 1 GeV), we observe that the running of the coupling freezes, indicating the breakdown of matching to 4-loop perturbation theory.
We convert our results for the coupling
to $\Lambda_{\MSb}^{N_f=3}$ \cite{Callan:1970yg,Symanzik:1970rt} separately at each distance, see Fig.~\ref{fig: final}.
We show the perturbative running of $\alpha_s$ using our final value of the $\Lambda$-parameter in Fig.~\ref{fig: alpha results} and we discuss it below, after addressing systematic effects in our determination.

\noindent\textit{Final result}:
We consider several sources of uncertainty in our analysis and we decompose the error of our final result for the $\Lambda$-parameter according to these different sources.
The raw lattice correlators are, obviously, subject to statistical errors (``lat stat'').
Extrapolating the correlators to the chiral limit has its associated systematic uncertainty (``chiral'').
The perturbative subtraction of discretization effects via NSPT is also subject to statistical errors (``NSPT stat'') and moreover, to a systematic uncertainty of extrapolation of NSPT results to the infinite volume limit (``NSPT infV'').
The latter is computed as the difference between a polynomial fit to several volumes ranging from $V=32^4$ up to $V=80^4$ and the estimates from the largest volume $V=80^4$.
The correlator interpolation uncertainty, described above, is denoted by ``interpol''.
Renormalizing the correlators in the $\MSb$ scheme introduces an uncertainty from the values of $Z$-factors (``$Z_A$'' and ``$Z_V$''). The uncertainty of the $c_V$ and $c_A$ improvement coefficients is completely negligible compared to its other sources.
Finally, we estimate the truncation uncertainty of the final $\Lambda$-value as the difference between conversions of $\alpha_s$ results to $\Lambda$ using the 4-loop and 5-loop $\beta$-functions (``trunc'').
These differences are shown in Fig.~\ref{fig: final}, including also the 2-loop and 3-loop cases.
The observed behavior suggests that while 3-loop perturbation theory still shows significant truncation effects in the considered energy range, the 4-loop and 5-loop results evince convergence. We double this uncertainty to cover the truncation of the perturbative series of Eq.~(\ref{eq. strategy}), where the 5-loop coefficient is not available at present.  

To make our final result independent from the choice of the window of physical distances where $\alpha_s$ is extracted, we adopt a systematic procedure similar to the one used in Ref.~\cite{Cichy:2016qpu}.
From all distances smaller than 0.2 fm, above which the coupling freezes, we choose the range $0.13$-$0.19$ fm, where all other systematic uncertainties are under good control.
Having 7 determinations of $\Lambda$ corresponding to these different distances, we calculate all possible weighted averages covering from one to seven subsequent distances.
We use the 28 resulting values of $\Lambda$ to build a weighted histogram, where the weights are taken as the squared inverse error of each individual result.
The histogram is approximately Gaussian (see the inset of Fig.~\ref{fig: final}) and we fit its mean and width to determine the central value, i.e.\ $\Lambda_{\MSb}^{N_f=3}$, and its uncertainty from the choice of the physical distances (``window'').
This central value, along with the total uncertainty, is shown as the green band in Fig.~\ref{fig: final}. 

The final result for the $\Lambda$-parameter reads:
\begin{eqnarray}
\Lambda_{\MSb}^{N_f=3} &=& 342(2.9)^{\rm lat}_{\rm stat}(5.0)_{\rm chiral}(6.5)^{\rm NSPT}_{\rm stat}(6.4)^{\rm NSPT}_{\rm infV}(0.8)_{Z_A} \nonumber \\
&\,&(1.0)_{Z_V} (0.4)_{\rm interpol}(4.8)_{\rm trunc} (12)_{\rm window} \textrm{\,MeV} \nonumber \\
&=& 342 (17) \textrm{\,MeV},
\end{eqnarray}
where we combine the individual uncertainties in quadrature to obtain the total error.
Our final value agrees well with earlier lattice determinations, e.g.\ with the recent one of Ref.~\cite{Bruno:2017gxd}, $\Lambda_{\MSb}^{N_f=3}=341(12) \textrm{\,MeV}$, and with a comparable total error, dominated in our case by the uncertainty from the choice of the physical distance and by the uncertainty from the NSPT correction.

\noindent\textit{Discussion and Conclusions}: 
In this Letter, we presented and tested a novel method to estimate the $\MSb$ strong coupling constant using numerical simulations of coordinate-space correlators and used it to determine the 3-flavor QCD $\Lambda$-parameter. It is based on current-current correlation functions in position space at small distances.
Our results suggest that the challenging multiscale problem of evaluating $\alpha_s$ can be addressed using lattices available today. 
We have shown that using a combination of state-of-the-art simulations and novel analysis techniques, one can find a window of available scales $\mu$ and provide an estimate of $\Lambda_{\MSb}^{N_f=3}$ with a competitive precision. In particular, the crucial steps are the perturbative subtraction of hypercubic artifacts and a combined continuum extrapolation using four lattice spacings and several lattice directions, which allowed us to control discretization effects at small distances in lattice units. 
We, furthermore, profited from independent evaluations of axial and vector correlators, which have a common continuum limit at short distances, to design a criterion to characterize the quality of continuum extrapolations and gain confidence in the results.

To conclude, we believe that the techniques described in this Letter provide a robust way of extracting the running of the QCD coupling and the QCD $\Lambda$-parameter, with good statistical precision and well-controlled sources of systematic effects.
Furthermore, the precision reached in this work can be increased even more in a systematic way.
Techniques based on current-current correlators in position space, improved by NSPT reduction of discretization effects, can be useful to determine other quantities, such as the quark condensate.

\section{Acknowledgments}
We gratefully acknowledge discussions with V.~Braun, F.~Knechtli and T.~Korzec. 

This research was carried out with the support of the Interdisciplinary Centre for Mathematical and Computational
Modelling (ICM) University of Warsaw under grant No. GA67-12, GA69-20, GA71-26, GA76-14 and AGH Cyfronet Computing Center under grant No. pionda, nspt, hadronspectrum. This work was supported by Deutsche Forschungsgemeinschaft under Grant No. SFB/TRR 55
and in part by the polish NCN grants No. 2016/21/B/ST2/01492 (P.K. and S.C.) and 2016/22/E/ST2/00013 (K.C.). P.K. acknowledges support from the NAWA Bekker fellowship  and  thanks  Universit\`{a}  degli Studi  di  Roma  ”Tor  Vergata”  for hospitality during which this work has been initiated.

We  thank  our  colleagues  in  the  Coordinated  Lattice  Simulations  (CLS)  effort  [http://wiki-zeuthen.desy.de/CLS/CLS] for the joint generation of the gauge field ensembles on which the computation described here is  based.
The gauge ensembles were generated with the help of the Gauss Centre for
Supercomputing e.V. (http://www.gauss-centre.eu) using computer time allocations 
on SuperMUC at Leibniz Supercomputing Centre (LRZ, http://www.lrz.de) and
JUQUEEN at Jülich Supercomputing Center (JSC, http://
www.fz-juelich.de/ias/jsc).
GCS is the alliance of the three national supercomputing centers HLRS (Universität
Stuttgart), JSC (Forschungszentrum Jülich) and LRZ
(Bayerische Akademie der Wissenschaften), funded by
the German Federal Ministry of Education and Research
(BMBF) and the German State Ministries for Research of
Baden-Württemberg (MWK), Bayern (StMWFK) and
Nordrhein-Westfalen (MIWF). 
Additionally computer time provided by PRACE
(Partnership for Advanced Computing in Europe, http://
www.prace-ri.eu) as part of the project ContQCD was used.
Additional simulations were performed on the
Regensburg iDataCool cluster and on the SFB/TRR 55
QPACE computer \cite{Baier:2009yq}, \cite{NAKAMURA2011841}. OPENQCD \cite{Luscher2013519} was used to
generate the main gauge ensembles, as part of the joint CLS
effort \cite{Bruno:2014jqa}. Additional ensembles were generated
on QPACE (using BQCD \cite{NAKAMURA2011841},\cite{Hoelbling:2011kk}) and on the Wilson HPC
Cluster at IKP Mainz.
\onecolumngrid
\appendix
\renewcommand{\theequation}{A-\arabic{equation}}
\setcounter{equation}{0}  
\renewcommand\thefigure{A-\arabic{figure}}    
\setcounter{figure}{0}    
\section{APPENDIX A: SUPPLEMENTAL MATERIAL}
\section{Chiral limit of correlation functions}
\label{sec:chiral}
The matching of lattice-extracted correlation functions to perturbation theory is performed in the massless $\MSb$ scheme.
Thus, all correlators obtained at non-zero quark masses need to be extrapolated to the chiral limit.
We use two kinds of a fitting ansatz for the chiral extrapolation for each lattice point:
\begin{equation}
\label{eq:chiral1}
C_{\Gamma}(1/x,m_q,a)=C_{\Gamma}(1/x,a)+\alpha_{m_q}y 
\end{equation}
and
\begin{equation}
\label{eq:chiral2}
C_{\Gamma}(1/x,m_q,a)=C_{\Gamma}(1/x,a)+\alpha_{m_q}y+\beta_{m_q}y^2,
\end{equation}
where $y = t_0 m^2_{\pi}$ is dimensionless and $t_0$ is an intermediate unphysical scale introduced in Ref.~\cite{Luscher:2010iy}.
We use values of $t_0/a^2$ from Ref.~\cite{inprep2}.
In both fitting ansatzes, the fitting parameter corresponding to the chiral limit value is denoted by $C_{\Gamma}(1/x,a)$ and the coefficients $\alpha_{m_q}$ and $\beta_{m_q}$ describe, respectively, effects linear and quadratic in $y$.

\begin{figure}
\includegraphics[width=0.49\textwidth]{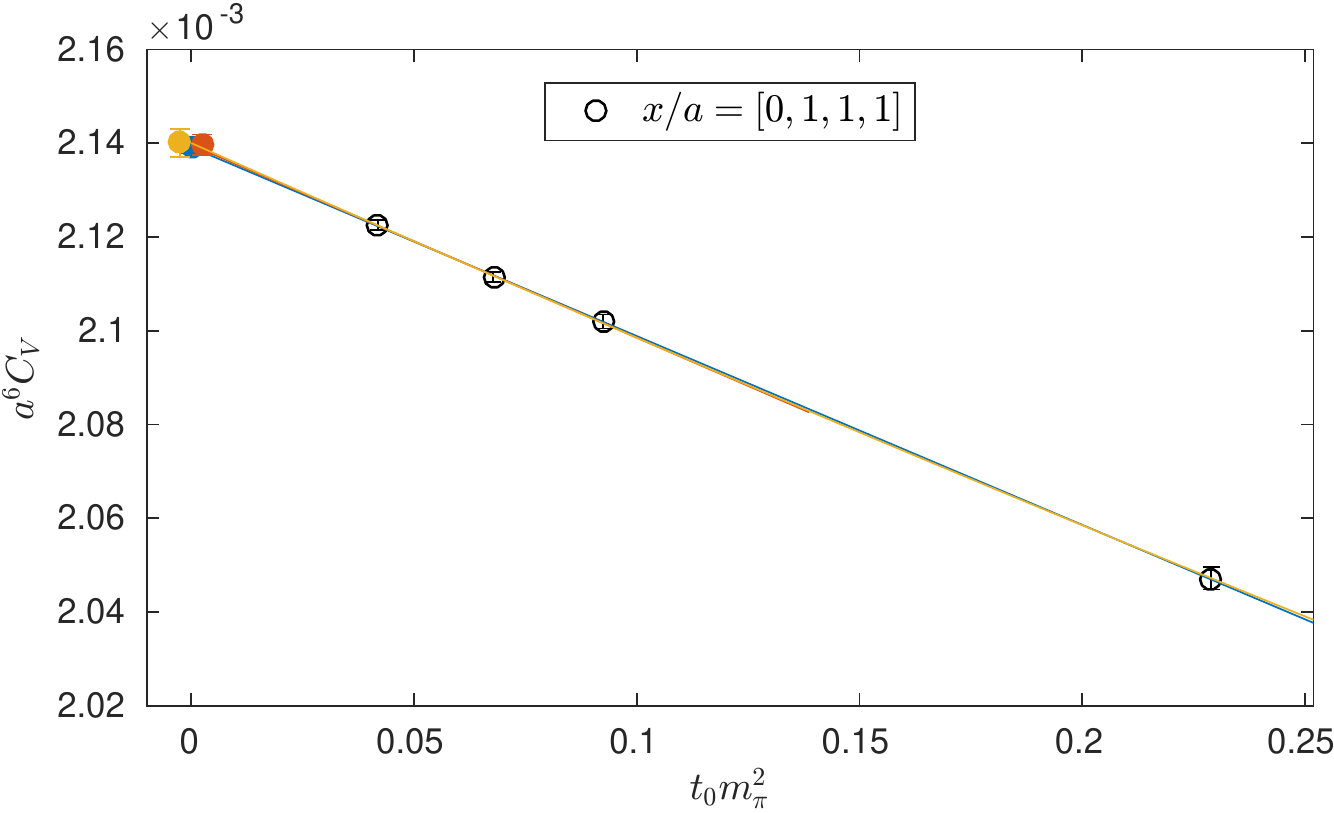}
\includegraphics[width=0.49\textwidth]{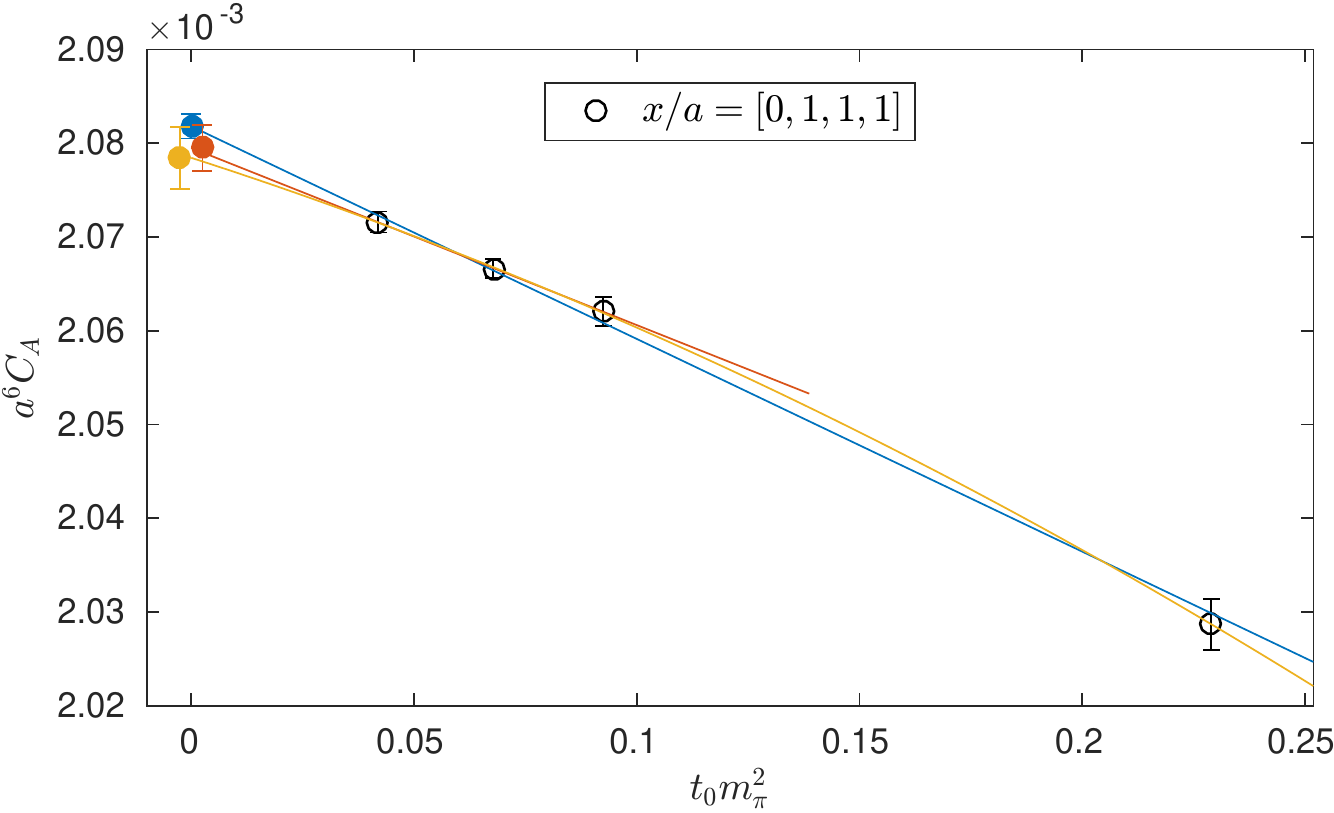}
\includegraphics[width=0.49\textwidth]{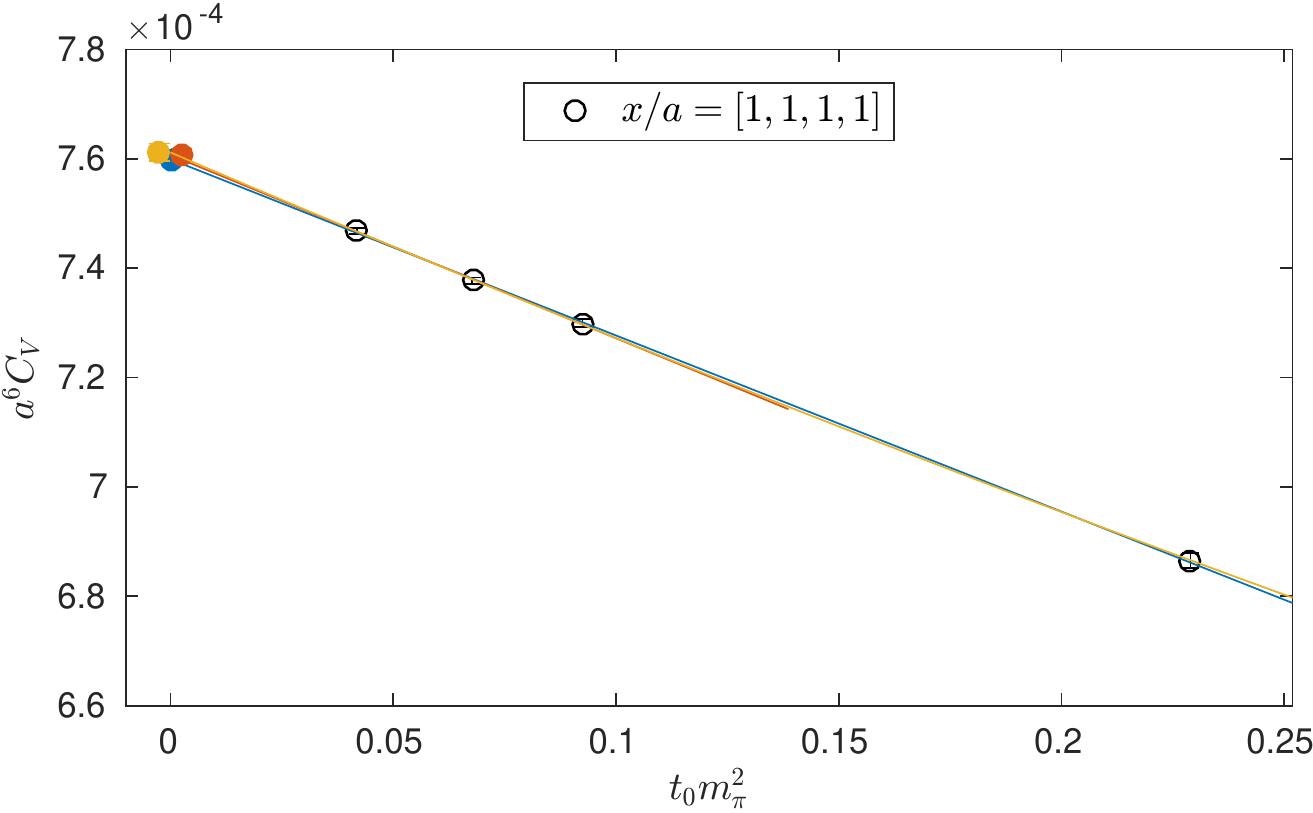}
\includegraphics[width=0.49\textwidth]{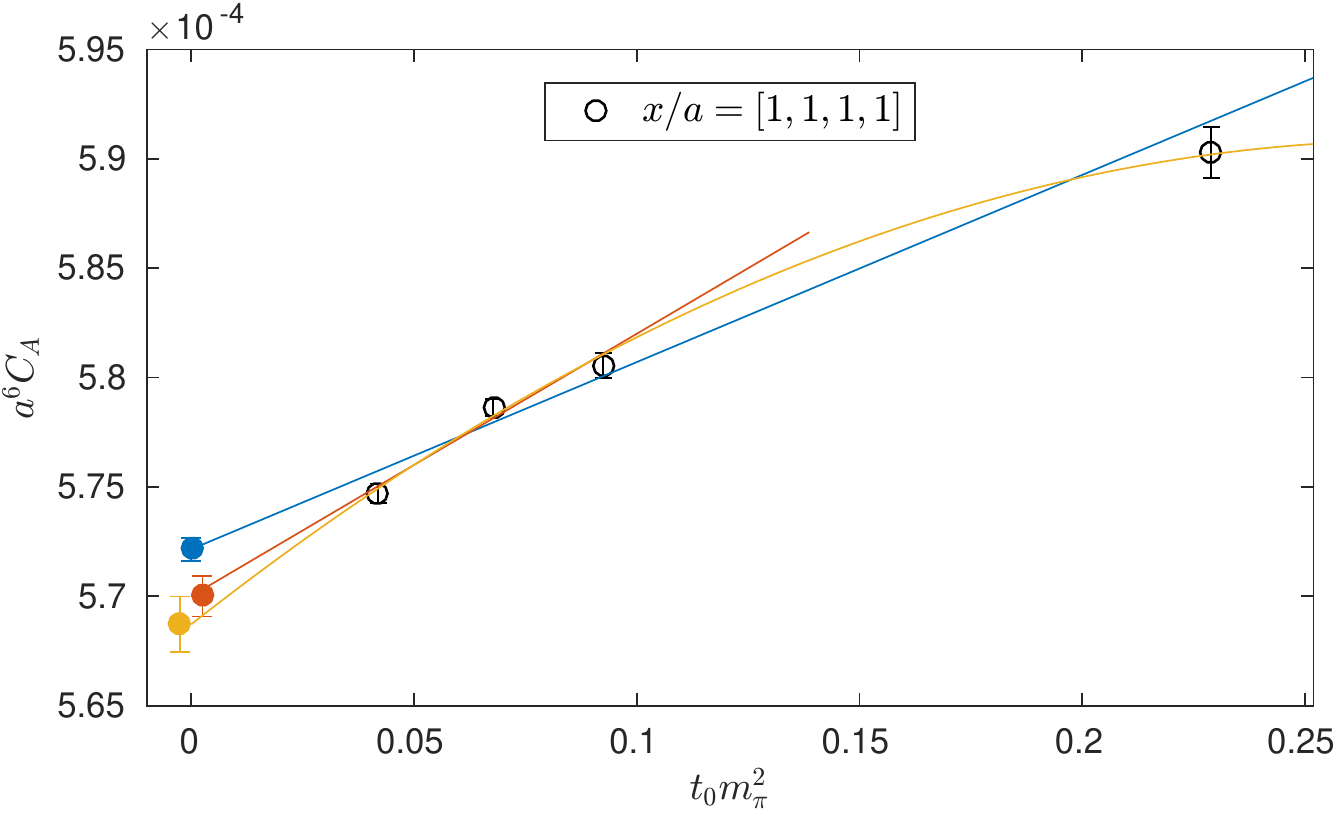}
\includegraphics[width=0.49\textwidth]{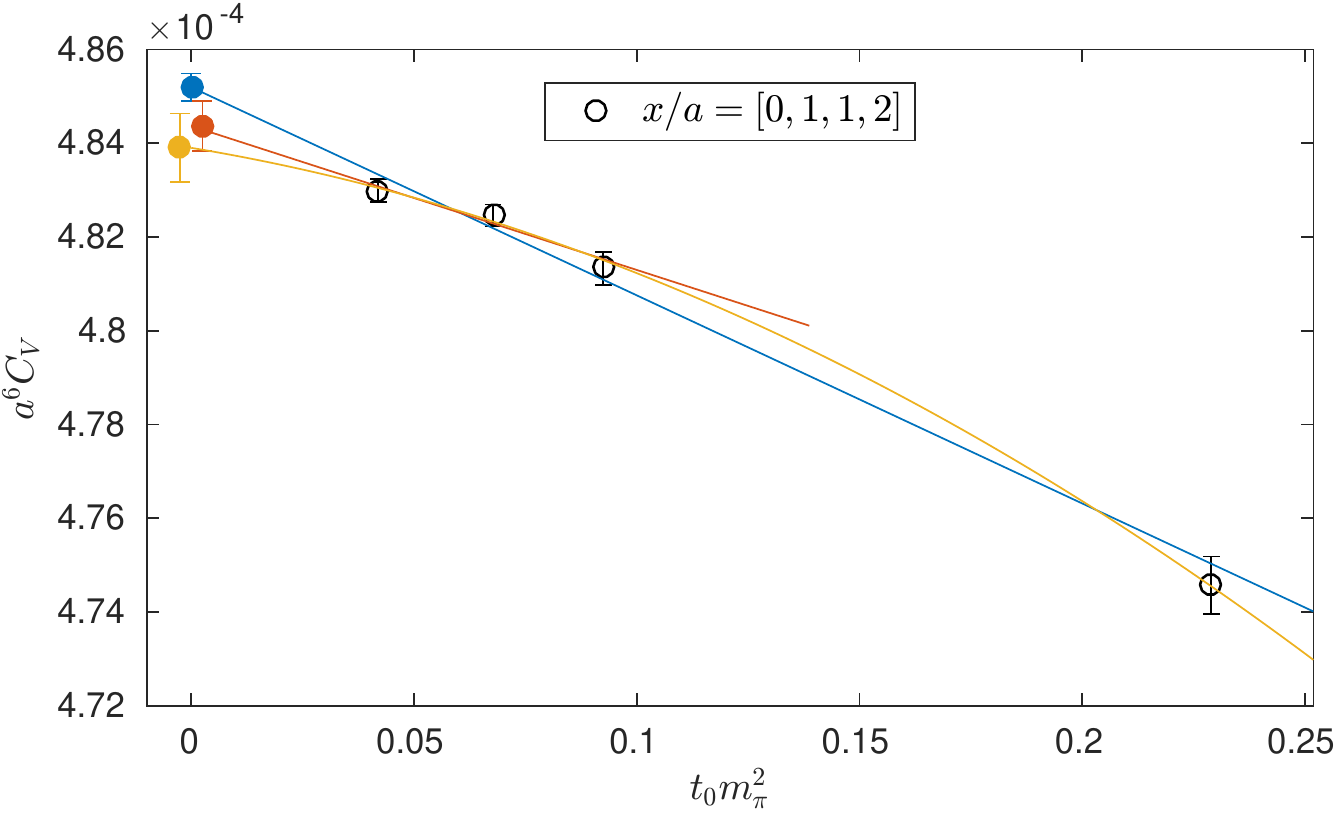}
\includegraphics[width=0.49\textwidth]{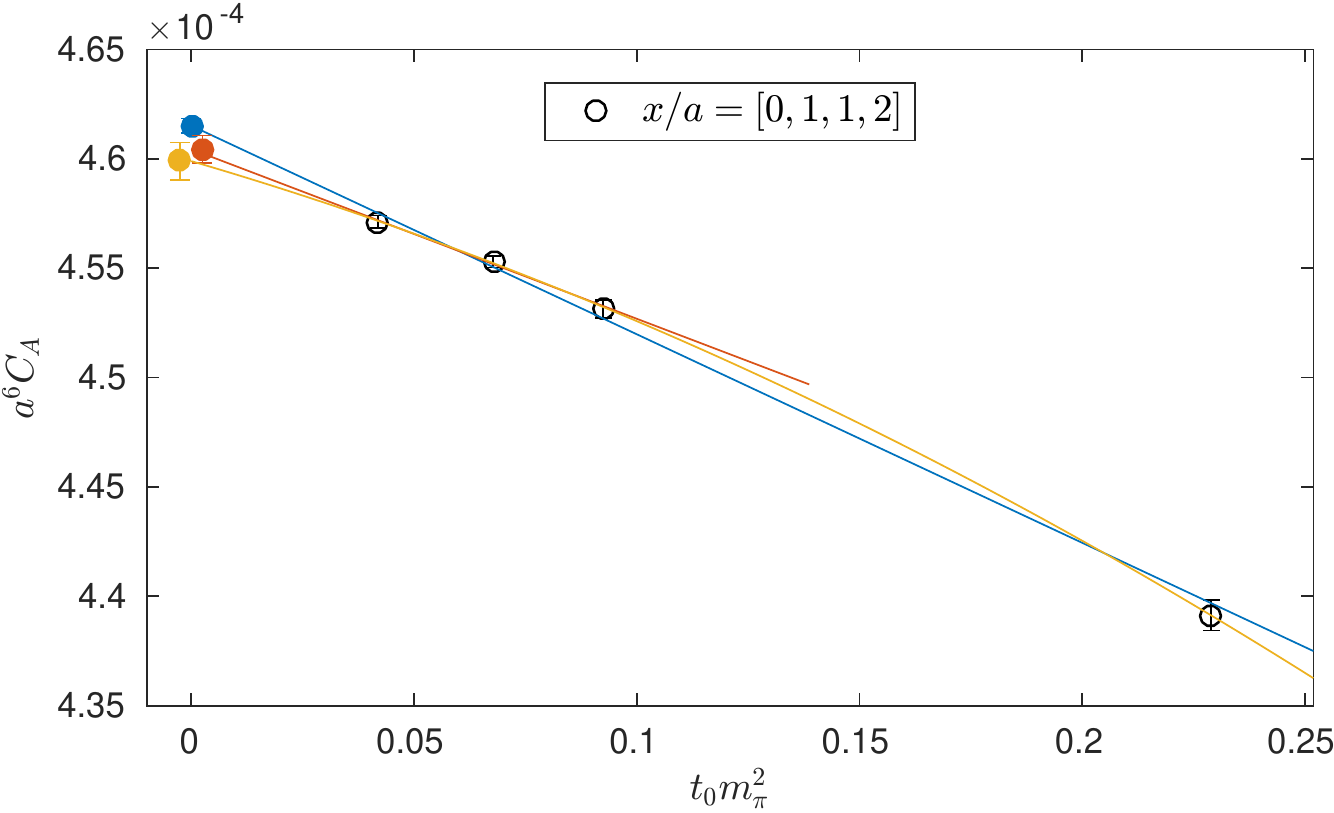}
\caption{Chiral limit extrapolations of the vector (left) and axial vector (right) correlator at $\beta=3.55$, where we have 4 pion masses available (around 270, 350, 410 and 710 MeV). Shown are linear fits to 3 or 4 masses and quadratic fits to 4 masses. \label{fig:chiral}}
\end{figure}

We tested both chiral extrapolations at $\beta=3.55$.
At this lattice spacing, we have four ensembles differing in the quark mass, corresponding to pion masses of around 270, 350, 410 and 710 MeV.
We performed linear fits in $y$ using 3 lightest pion masses or all 4 masses and compared them to quadratic fits employing all masses.
Our fits are shown in Fig.~\ref{fig:chiral} for both vector and axial vector correlators and for all 3 types of points that we use in our analysis (see below for the discussion of this choice) -- $(0,k,k,k)$, $(k,k,k,k)$ and $(0,k,k,2k)$ (for $k=1$).
In 3 cases, we observe full agreement between the 3 kinds of fits.
The linear fit to pion masses up to 410 MeV always agrees with the quadratic one including the 710 MeV point.
Tension is observed between the linear and quadratic fit to all masses in 3 cases.
However, to a large extent it is a consequence of the excellent statistical precision of the data.
This is concluded when looking at the relative deviation of the chiral limit value from the quadratic fit and from the linear fit.
This deviation reads $0.04\%$, $0.16\%$, $0.16\%$, $0.60\%$, $0.27\%$ and $0.35\%$ (in order of plots in Fig.~\ref{fig:chiral}).
We also quote (smaller) deviations from comparing the linear fits to 3 or 4 masses: $0.03\%$, $0.11\%$, $0.11\%$, $0.38\%$, $0.17\%$ and $0.23\%$.

Since we cannot to perform such an analysis at other values of $\beta$ with the available ensembles of gauge field configurations, we estimate the uncertainty from the chiral limit extrapolation in the following way.
We consider the average deviation between the linear and quadratic fit to all 4 masses and between the linear fits to 3 and 4 masses.
The former is $0.26\%$ and the latter $0.17\%$.
To be conservative, we take the larger number as our estimate of the chiral limit uncertainty at the level of correlators.
Then, we run a bootstrap procedure including all our $\beta$ values and we find the corresponding uncertainty at the level of $\alpha_s$ and, finally, of the $\Lambda$-parameter.
The implied uncertainty at the level of the latter is 5 MeV, i.e.\ it is comparatively larger than for the correlation functions.
However, it is still subleading with respect to other uncertainties present in our analysis.

We also note that the small magnitude of mass effects in coordinate-space correlators seems to hold more generally, as they were found to be subleading also in our previous analyses aimed at extracting renormalization functions of local bilinear operators \cite{Cichy:2012is} and the running of the quark mass \cite{Cichy:2016qpu}.
\section{Perturbative subtraction of hypercubic artefacts}
\label{sec:NSPT}
\subsection{Theoretical principles}
Correlation functions that are the basis of this work are computed on the lattice for several points $X$, corresponding to different energy scales related to the distance $x=\sqrt{X^2}$.
The lattice breaks the continuum rotational symmetry and thus, correlators evaluated at the same distance, but using points inequivalent with respect to the remaining hypercubic symmetry, may be significantly different.
In other words, the breaking of rotational symmetry by the lattice induces hypercubic artefacts that contaminate the correlators.
Even though such artefacts vanish in the continuum limit, the induced contamination is, in practice, severe and prevents meaningful extraction of $\alpha_s$ from this approach.
The problem is less significant when extracting renormalization functions of local bilinear operators, as done in Refs.~\cite{Cichy:2012is,Cichy:2016qpu}.
In these works, a tree-level subtraction of hypercubic artefacts was employed and combined with ``democratic cuts'', to be defined below.

The tree-level-corrected correlator in the chiral limit, $C^{[0]}(1/x,a)$, is defined as:
\begin{equation}
\label{eq:tlcorr}
C_{\Gamma}^{[0]}(1/x,a)=C_{\Gamma}(1/x,a)+C_{\Gamma}^{\rm free,\,cont}(1/x)-C_{\Gamma}^{\rm free,\,lat}(a/x),
\end{equation}
where $C_{\Gamma}^{\rm free,\,cont}(1/x)=6/(\pi^4x^6)$ is the free continuum correlator (equal for the vector and axial vector cases) and $C_{\Gamma}^{\rm free,\,lat}(a/x)$ is the free lattice correlator (computed with the same fermionic discretization as the interacting correlator).
Thus, the tree-level correction subtracts the discretization effects appearing in the non-interacting case.

For the one-loop correction, we proceed in the following way. We use the perturbative continuum expression for the current-current correlation function in position space renormalized in the $\MSb$ scheme at the scale $\mu=1/x$ from Ref.~\cite{Chetyrkin:2010dx}, which we summarize schematically as
\begin{equation}
    C_{\Gamma}^{\MSb,\,\cont}(1/x,g_0^2) = C_{\Gamma}^{\rm free,\,cont}(1/x) + g_0^2 C_{\Gamma}^{\rm 1-loop,\, cont}(1/x),
\end{equation}
where the index '1-loop' indicates the isolated 1-loop contribution. $g_0$ is the bare coupling constant.
In parallel, we have the lattice evaluation of the same renormalized quantity, $C_{\Gamma}^{\MSb,\,\lat}(a/x,g_0^2)$, for which we employ numerical stochastic perturbation theory (NSPT; details of the calculations are described in the following section). We again summarize them schematically as
\begin{equation}
    C_{\Gamma}^{\MSb,\,\lat}(a/x,g^2_0) = \left(Z_{\Gamma}^{\MSb}(g_0^2)\right)^2 \left( C^{\rm free,\,\lat}(a/x) + g_0^2 C^{\rm 1-loop,\,\lat}(a/x) \right),
\end{equation}
where the renormalization factors for the axial and vector currents for the relevant action were evaluated in lattice perturbation theory (LPT) \cite{Taniguchi:1998pf},
\begin{align}
    \left(Z^{\MSb}_V(g_0^2)\right)^2 &= 1 + 2 g_0^2 (-0.10057) \equiv 1 + 2 g_0^2 Z^{\textrm{1-loop}}_V, \label{eq.renfactor V} \\
    \left(Z^{\MSb}_A(g_0^2)\right)^2 &= 1 + 2 g_0^2 (-0.09048) \equiv 1 + 2 g_0^2 Z^{\textrm{1-loop}}_A.
    \label{eq.renfactor A}
\end{align}
The continuum, $C_{\Gamma}^{\MSb,\,\cont}(1/x,g_0^2)$, and lattice, $C_{\Gamma}^{\MSb,\,\lat}(a/x,g_0^2)$, correlators differ by discretization effects. Hence, we can use both of them to improve our non-perturbative data as follows,
\begin{align}
    C_{\Gamma}^{[1]}(1/x,a,g_0^2) &= C_{\Gamma}(1/x,a) + C_{\Gamma}^{\MSb,\,\cont}(1/x,g_0^2) - C_{\Gamma}^{\MSb,\,\lat}(a/x,g_0^2) \nonumber \\
    &= C_{\Gamma}(1/x,a) + 
    C_{\Gamma}^{\rm free,\,cont}(1/x) + g_0^2 C_{\Gamma}^{\rm 1-loop,\, cont}(1/x) \nonumber\\
    &\qquad\qquad\qquad-\left(Z_{\Gamma}^{\MSb}(g_0^2)\right)^2 \left( C^{\rm free,\,\lat}(a/x) + g_0^2 C^{\rm 1-loop,\,\lat}(a/x) \right)
    \Big).
\end{align}
Finally, $C^{[1]}(1/x,a,g_0^2)$ is the 1-loop corrected lattice correlation function, obtained by applying both the tree-level and the 1-loop NSPT correction to the non-perturbative (chirally-extrapolated, see previous section) lattice data, $C_{\Gamma}(1/x,a)$.

Correlators computed at different lattice points differ in the degree to which the continuum rotational symmetry is broken.
Thus, they differ in the extent the one-loop correction can restore this symmetry, i.e.\ subtract hypercubic artefacts.
The extent of symmetry breaking is related to the value of the invariant
\begin{equation}
X^{[4]}\equiv\frac{(\sum_\mu X_\mu^4)}{(\sum_\mu X_\mu^2)^2}, 
\label{X4}
\end{equation}
i.e.\ points with smaller $X^{[4]}$ tend to evince smaller hypercubic artefacts than points with the same $X^2$, but larger $X^{[4]}$.
Another proxy for the size of these artefacts is the slope of the line going through the given point and the origin with respect to the line containing the point $(1,1,1,1)$.
We denote such an angle by $\theta$.

In perturbative subtraction of hypercubic artefacts in renormalization functions extracted in momentum space, the analogous invariant is $P^{[4]}$.
It was found in Ref.~\cite{Alexandrou:2015sea} that the 1-loop correction works best for points with $P^{[4]}\lesssim0.4$, see e.g.\ Fig.\ 15 of this reference for the effects in the renormalization factor of the axial vector current, $Z_A$.
The points with $P^{[4]}>0.4$ also have a large part of the artefacts subtracted, but they do not fall onto a universal curve formed by the corrected points with $P^{[4]}<0.4$.
The criterion with the angle $\theta$ was, in turn, used in the coordinate-space extraction of $Z$-factors \cite{Cichy:2012is,Cichy:2016qpu}.
As shown e.g.\ in Fig.~4 of Ref.~\cite{Cichy:2012is} for the renormalization function of the pseudoscalar density, $Z_P$, applying only the tree-level correction is not enough to obtain smooth dependence of the $Z$-factor on the distance, whereas restricting to points with $\theta\leq30$ degrees already allows to include points that fall onto a universal plateau.
Note, however, that even though a plateau is observed within statistical uncertainties, there are remaining trends in the behavior of points, indicating that they have not been fully corrected.
These trends are the reason why meaningful extraction of $\alpha_s$ becomes unfeasible with only the tree-level correction with our present, statistically more precise data, due to a rather large sensitivity of the implied values of $\alpha_s$.
Hence, it necessitates the application of the 1-loop correction, never before applied to correlators in coordinate space.
Both criteria of $X^{[4]}\leq X^{[4]}_{\rm max}$ (or $P^{[4]}\leq P^{[4]}_{\rm max}$ in momentum space) and $\theta\leq\theta_{\rm max}$ are commonly referred to as ``democratic'' cuts and a proper choice of $X^{[4]}_{\rm max}$ or $\theta_{\rm max}$ needs to be made according to the statistical precision of the data.
We discuss this choice after describing our computation of the 1-loop NSPT correction and its effects on the lattice data.

\subsection{Calculation of the 1-loop correction in numerical stochastic perturbation theory}

\begin{figure}
\includegraphics[width=0.47\textwidth]{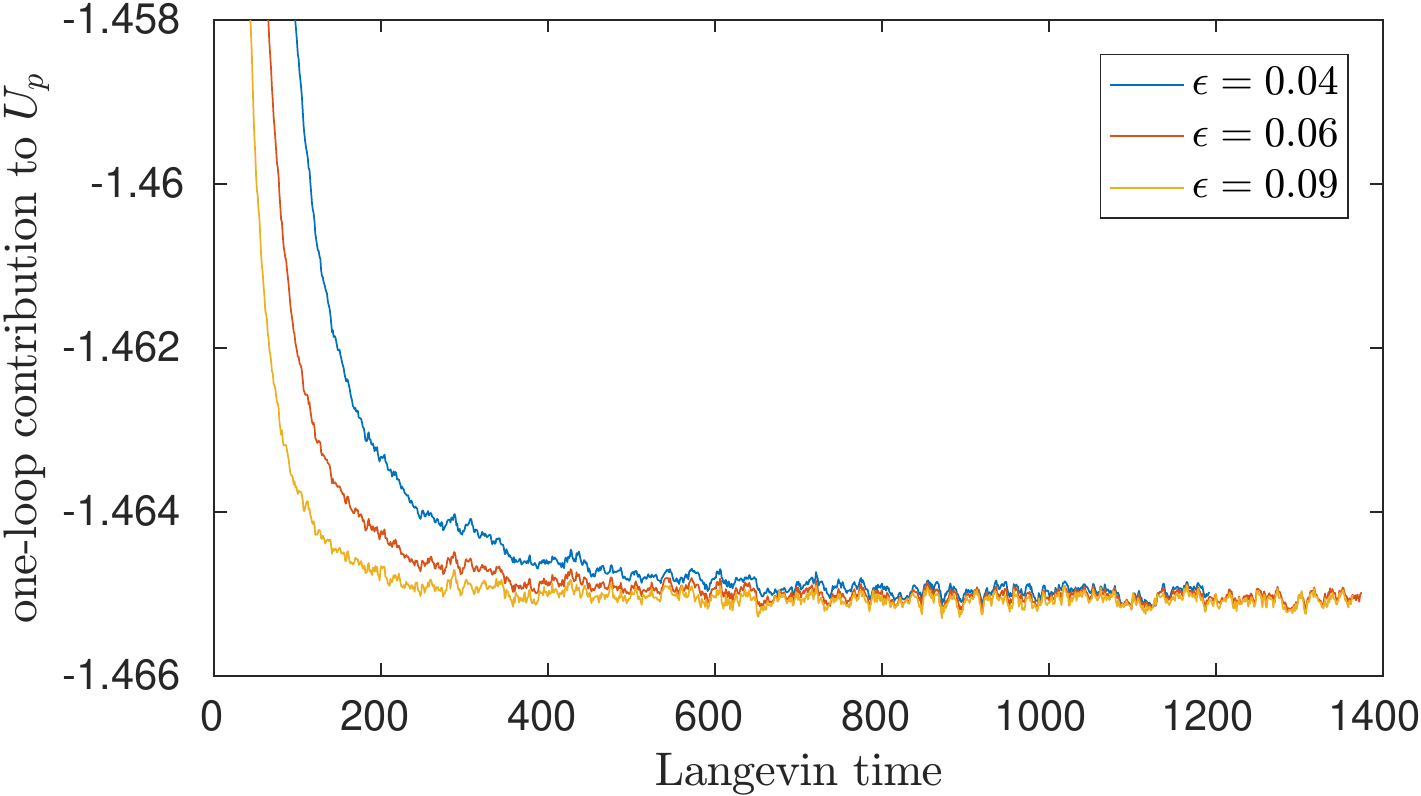}
\caption{Thermalization of Wilson plaquette for three simulations with different size of the discretized Langevin time. Larger step leads to a faster thermalization, hence the use of a quadratic integration scheme considerably reduces the needed computational effort. \label{fig:nspt_thermalization}}
\end{figure}

\begin{figure}
\includegraphics[width=0.47\textwidth]{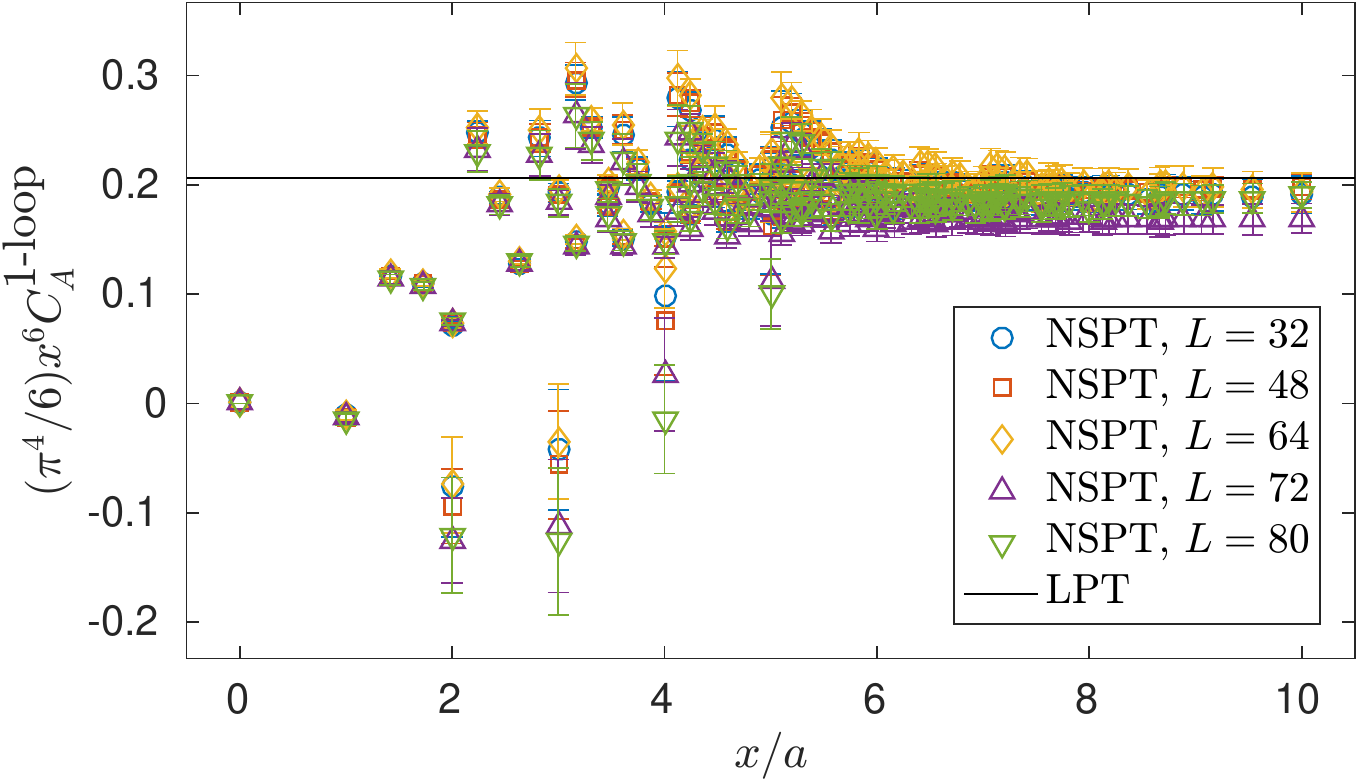}
\includegraphics[width=0.47\textwidth]{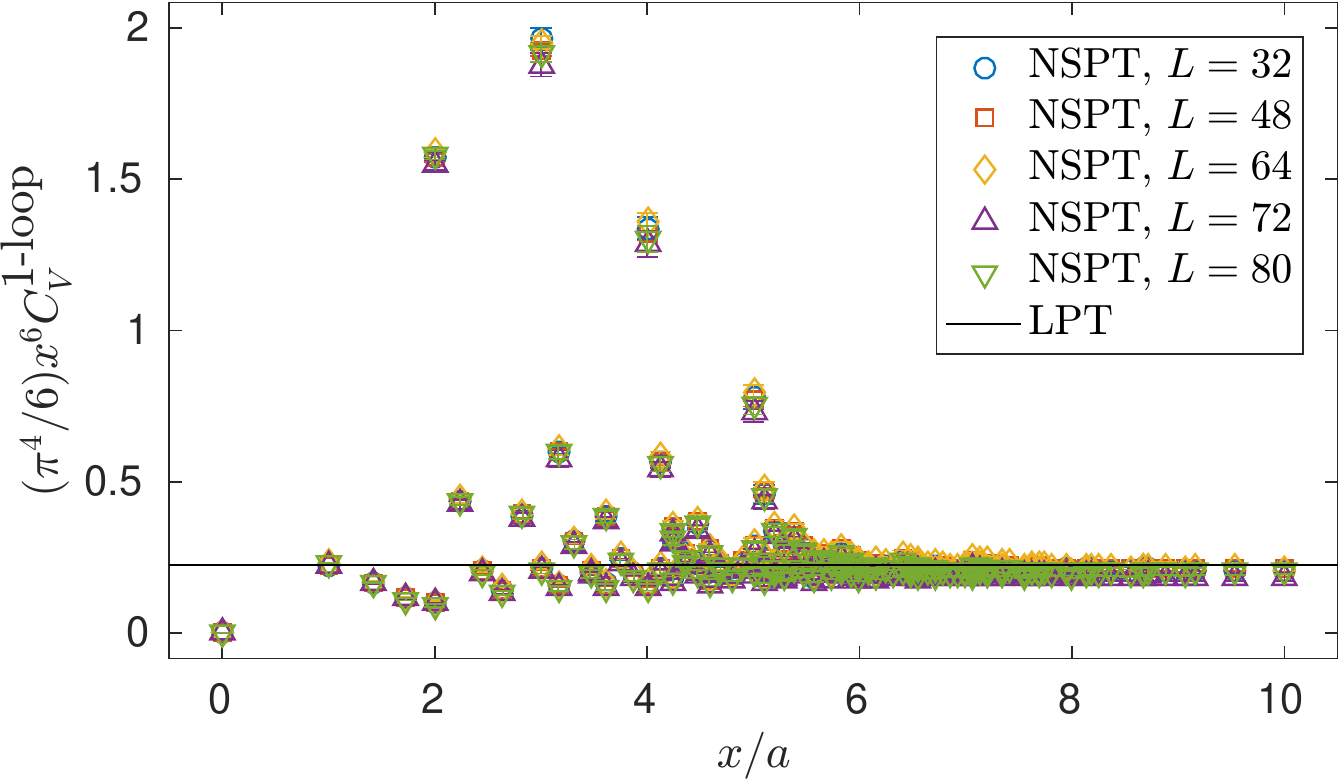}
\caption{Volume dependence of the one-loop contribution to the current-current correlators from NSPT. \label{fig:nspt_tests}}
\end{figure}

The framework of numerical stochastic perturbation theory (NSPT) was described in detail in Ref.~\cite{DiRenzo:2004hhl} and we follow this reference in our work. Using the Fortran code from Refs.~\cite{Simeth:2013ima,Simeth:2015xua}, we have implemented the L\"uscher-Weisz gluon action, as used in the nonperturbative CLS simulations \cite{Bruno:2014jqa}. The object of our studies are current-current correlators and the relevant next-to-leading order corrections to these correlators come only from gluon exchanges and hence, can be estimated using quenched NSPT ensembles. We performed multiple simulations with increasing volumes, $(L/a)^4$, ranging from $24^4$ up to $80^4$. The ensembles with volumes below $64^4$ were generated using the standard linear integration scheme, whereas for larger volumes we have implemented a second order integration scheme \cite{torrero2008determination,Bali_2013}, which significantly decreases the required amount of computer time by allowing simulations with larger Langevin step, see Fig.~\ref{fig:nspt_thermalization}.
In all cases, simulations for three values of the time discretization step were performed and the extrapolation to the continuum Langevin equation was performed linearly (smaller volumes) or quadratically (larger volumes). At the $g_0^2$ (1-loop) order, no problems with thermalization or ergodicity were noticed. In each ensemble, $\mathcal{O}(20)$ statistically independent configurations were generated, which turned out to be enough to have the statistical noise under control. On each configuration, the current-current correlation function was measured with the $g_0^0$ and $g_0^2$ orders being non-trivial,
\begin{equation}
C_{\Gamma}^{\rm NSPT}(a/x,g^2_0) = 
C_{\Gamma}^{\rm free,\,\lat}(a/x) + g_0^2 C_{\Gamma}^{\rm 1-loop,\,\lat}(a/x).
\label{eq. correlator1}
\end{equation}
We simulated massless fermions by using the known one-loop values of $\kappa_{\textrm{critical}}$ for the Wilson-clover discretization of the Dirac operator \cite{PhysRevD.65.014511}, which we have also reproduced from our numerical data as a check of the implementation of the clover term. The clover coefficient was set to 1. The correlation functions of Eq.~\eqref{eq. correlator1} were evaluated using 64 source positions spaced randomly over the volume of the configuration. We imposed $H_4$ invariance by averaging the correlation functions for all equivalent lattice distances $x$. We show the normalized data for the one-loop contribution to the axial-axial and vector-vector correlation functions for different volumes in Fig.~\ref{fig:nspt_tests} normalized to their tree-level values. We employed $\mathcal{O}(a)$ improved currents, with the massless improvement coefficients $c_V$ and $c_A$ set to their tree-level values. Note that the non-perturbative data was improved for the linear cut-off effects non-perturbatively, which ensures that we did not remove the $\mathcal{O}(a)$ term twice. 

\begin{figure}
\includegraphics[width=0.47\textwidth]{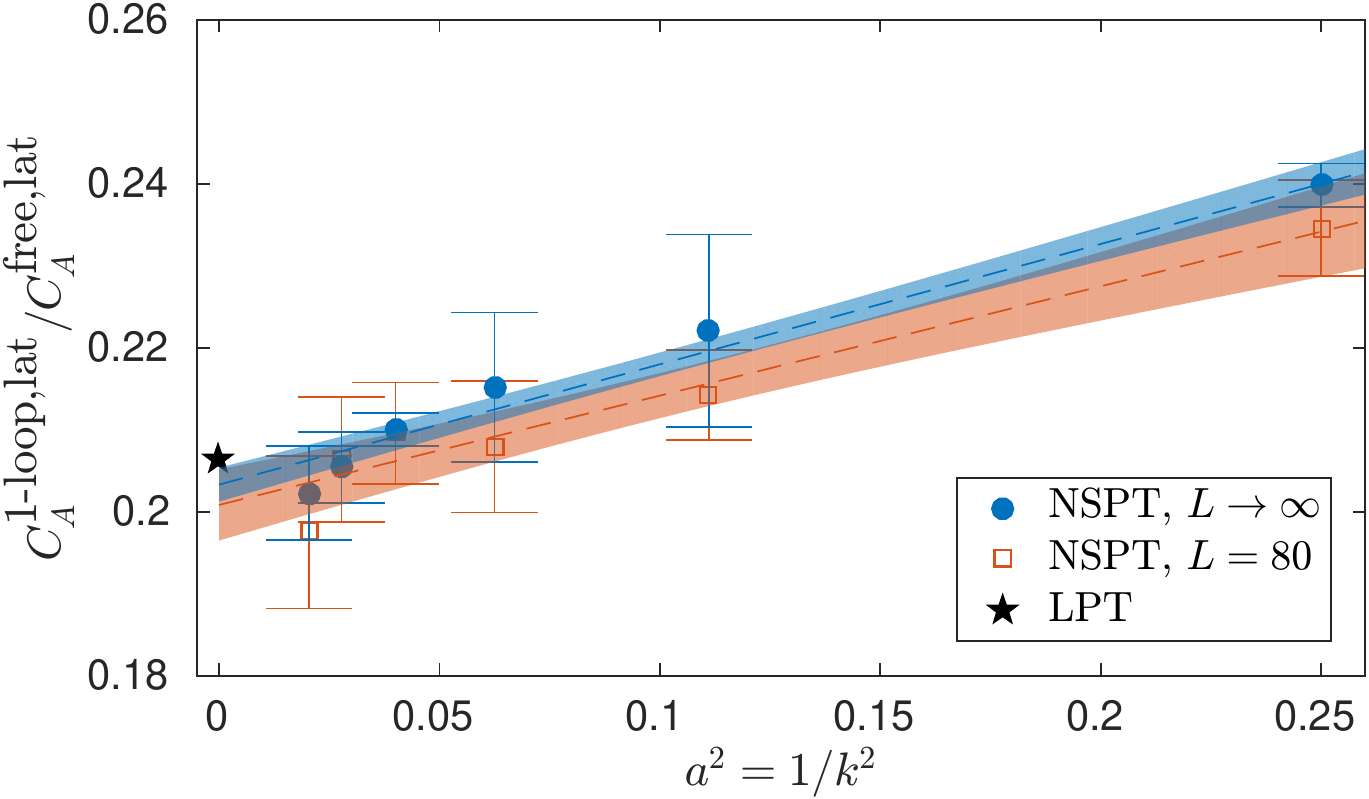}
\includegraphics[width=0.47\textwidth]{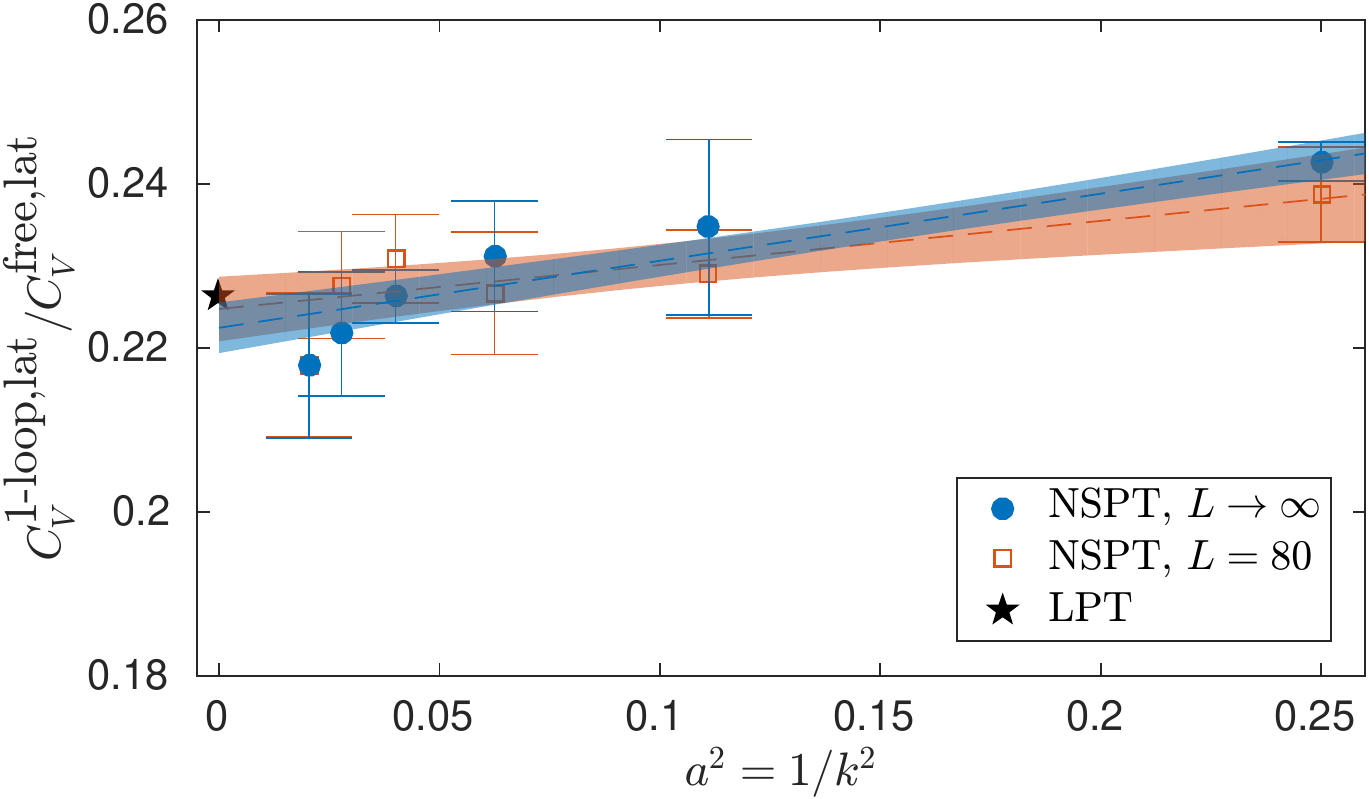}
\caption{Tests of the NSPT calculation. Comparison with the LPT calculation that yields an infinite volume value. The extrapolation is performed along a single lattice direction $(k,k,k,k)$ up to $k=7$ separately for the different volumes quadratically in the lattice spacing. Agreement with the LPT result confirms that finite volume effects are under control for the selected values of $k$ in this regime of volumes. \label{fig:nspt_tests_extrapolation}}
\end{figure}

As a test of the numerical setup, we used the known one-loop contributions to the axial and vector current renormalization factors already mentioned in Eqs.~\eqref{eq.renfactor V} and \eqref{eq.renfactor A} obtained in infinite volume from lattice perturbation theory. Using the numerical data for the correlation functions in position space, we can easily estimate the corresponding renormalization factors in the position space scheme  \cite{Chetyrkin:2010dx,Cichy:2012is}, $Z_{\Gamma}^X(a\mu,g_0^2)$, 
which we can perturbatively translate to the $\MSb$ scheme,
\begin{equation}
    \left(Z^{\MSb}_{\Gamma}(g_0^2)\right)^2 = \left(Z_{\Gamma}^X(a\mu, g_0^2)\right)^2 \left( 1 + \sum_{n>0} \delta_n^{\Gamma} \tilde{\alpha}_s^n(\mu) \right)^2 = \frac{C^{\rm free,\,\lat}_{\Gamma}(a/x)}{C^{\textrm{NSPT}}_{\Gamma}(a/x,g_0^2)} \left(1 + \sum_{n>0}\delta_n^{\Gamma} \tilde{\alpha}_s(\mu) \right)^2,
    \label{eq. ren test}
\end{equation}
where $a \mu = a/x$ and the numerical values of the perturbative conversion coefficients $\delta_n^{\Gamma}$ are given in Ref.~\cite{Chetyrkin:2010dx} to 4 loops and $\tilde{\alpha}_s(\mu)=\alpha_s(2e^{-\gamma_E}\mu)$ with $\alpha_s(\mu)$ being the $\MSb$ coupling (note the X-scheme $Z$-factors have a residual scale dependence from breaking of chiral Ward identities in this scheme \cite{Chetyrkin:2010dx}). 
Since we are interested in the one-loop contribution, we expand Eq.~\eqref{eq. ren test} and set $\tilde{\alpha}_s(\mu) = \frac{1}{4 \pi^2} g_0^2$ which is valid to the first order, 
\begin{equation}
    1 + 2 g_0^2 Z^{\textrm{1-loop}}_{\Gamma} = \frac{C^{\rm free,\,\lat}_{\Gamma}(a/x)}{C^{\rm free,\,\lat}_{\Gamma}(a/x) + g_0^2 C^{\textrm{1-loop},\,\lat}_{\Gamma}(a/x)} \left( 1 + 2 \frac{\delta_1^{\Gamma}}{4 \pi^2} g_0^2  \right).
    \label{eq. ren test 1 loop}
\end{equation}
Keeping only the $g_0^2$ terms gives
\begin{equation}
    2 Z^{\textrm{1-loop}}_{\Gamma} = -\frac{C^{\textrm{1-loop},\,\lat}_{\Gamma}(a/x)}{C^{\rm free,\,\lat}_{\Gamma}(a/x)}  + 2 \frac{\delta_1^{\Gamma}}{4 \pi^2}.
    \label{eq. ren test2}
\end{equation}
Hence, we can directly plot and compare the finite volume NSPT numerical data with the infinite volume LPT expectation,
\begin{equation}
     \frac{C^{\textrm{1-loop},\,\lat}_{\Gamma}(a/x)}{C^{\rm free,\,\lat}_{\Gamma}(a/x)} = - 2 Z^{\textrm{1-loop}}_{\Gamma} + 2 \frac{\delta_1^{\Gamma}}{4 \pi^2},
    \label{eq. ren test3}
\end{equation}
which we do in Fig.~\ref{fig:nspt_tests}. Clearly, at large distances where discretization effects are small, the data assume a plateau close to the expected value. With increasing volumes, the lattice values approach the expected continuum value marked by the solid line. The discrepancies should be associated to finite volume corrections. 
In order to check this, we show extrapolations to infinite distance in Fig.~\ref{fig:nspt_tests_extrapolation}. We use a subset of these data, namely data points along one lattice direction, $(k,k,k,k)$ for all available distances $k$ to perform the continuum extrapolation. The resulting fits for the largest employed volume and for the infinite volume limit are shown in Fig.~\ref{fig:nspt_tests_extrapolation}. Within statistical uncertainties, the continuum-extrapolated infinite-volume-limit value agrees both with the $L=80$ extrapolation and with the expectation from LPT.

\begin{figure}
\includegraphics[width=0.55\textwidth]{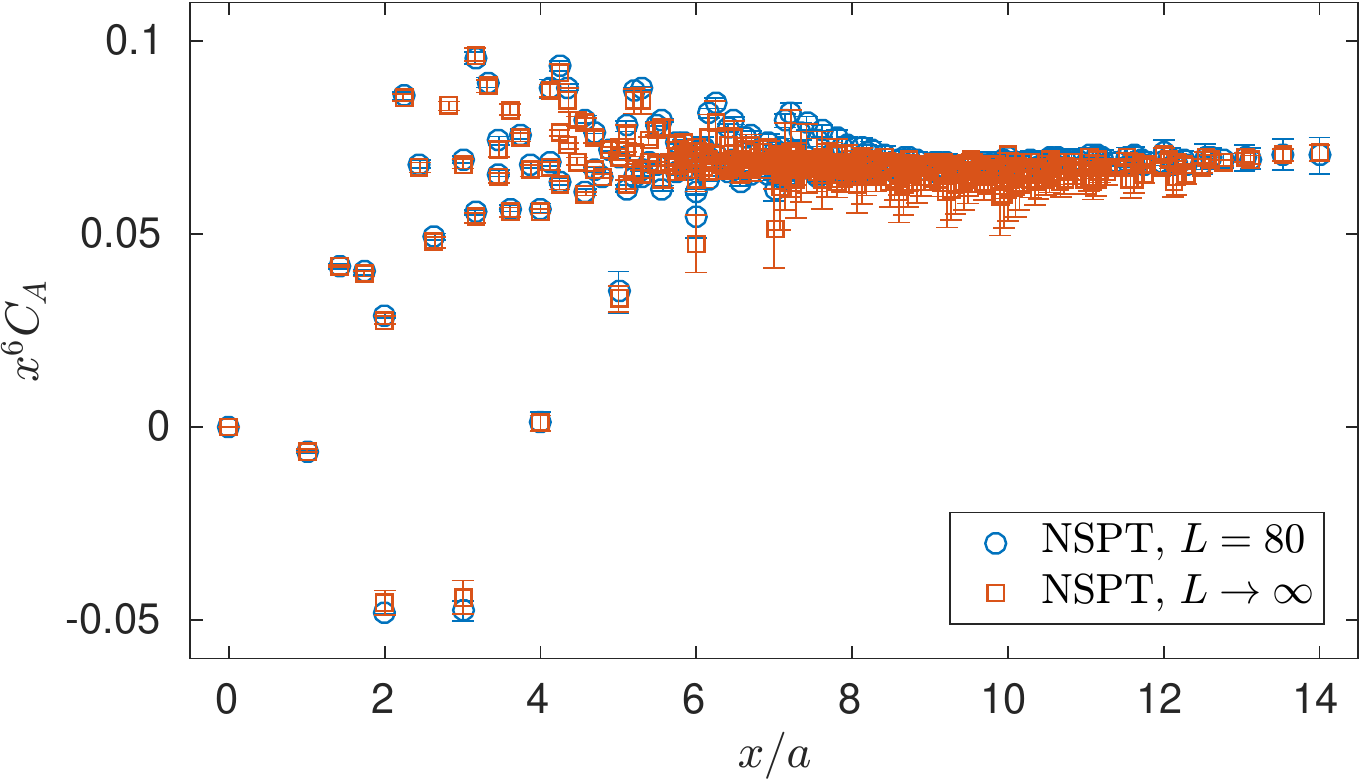}
\caption{Comparison of $L=80$ data with the infinite volume extrapolated values. \label{fig:nspt_extrapolation}}
\end{figure}

Nevertheless, we observe that there is some sensitivity to the lattice volume in our NSPT data, which could propagate to the non-perturbative data through the one-loop correction. In order to minimize this effect, we perform an infinite volume extrapolation of the one-loop correlation functions. We use the following fit ansatz,
\begin{equation}
 C_{\Gamma}^{\rm 1-loop,\,\lat}(a/x,L/a) = C_{\Gamma}^{\rm 1-loop,\,\lat}(a/x) + \left( c_1 x^2 + c_2 x^{[4]}/x^2 + c_3 x^{[6]}/x^{[4]} + c_4 x^{[8]}/x^{[6]}\right) / L^2,
 \end{equation}
which we use to perform a global fit to all volumes larger than $48^4$. 
The quantities $x^{[n]} = \sum_i x^n_i$ are invariants of the remaining $H_4$ symmetry group and we checked that adding more combinations of invariants did not lead to significantly different extrapolations, as long as the leading $x^2$ and one more term were included. As a final uncertainty of this step, we took a conservative estimate as the difference between the infinite volume estimate and the value for the $80^4$ lattice. This uncertainty combined quadratically with the statistical uncertainty of the NSPT correlation functions was taken as the final uncertainty of the one-loop correction. We show both the extrapolated and the $L=80$ data in Fig.~\ref{fig:nspt_extrapolation}. 

\begin{figure}
\includegraphics[width=0.49\textwidth]{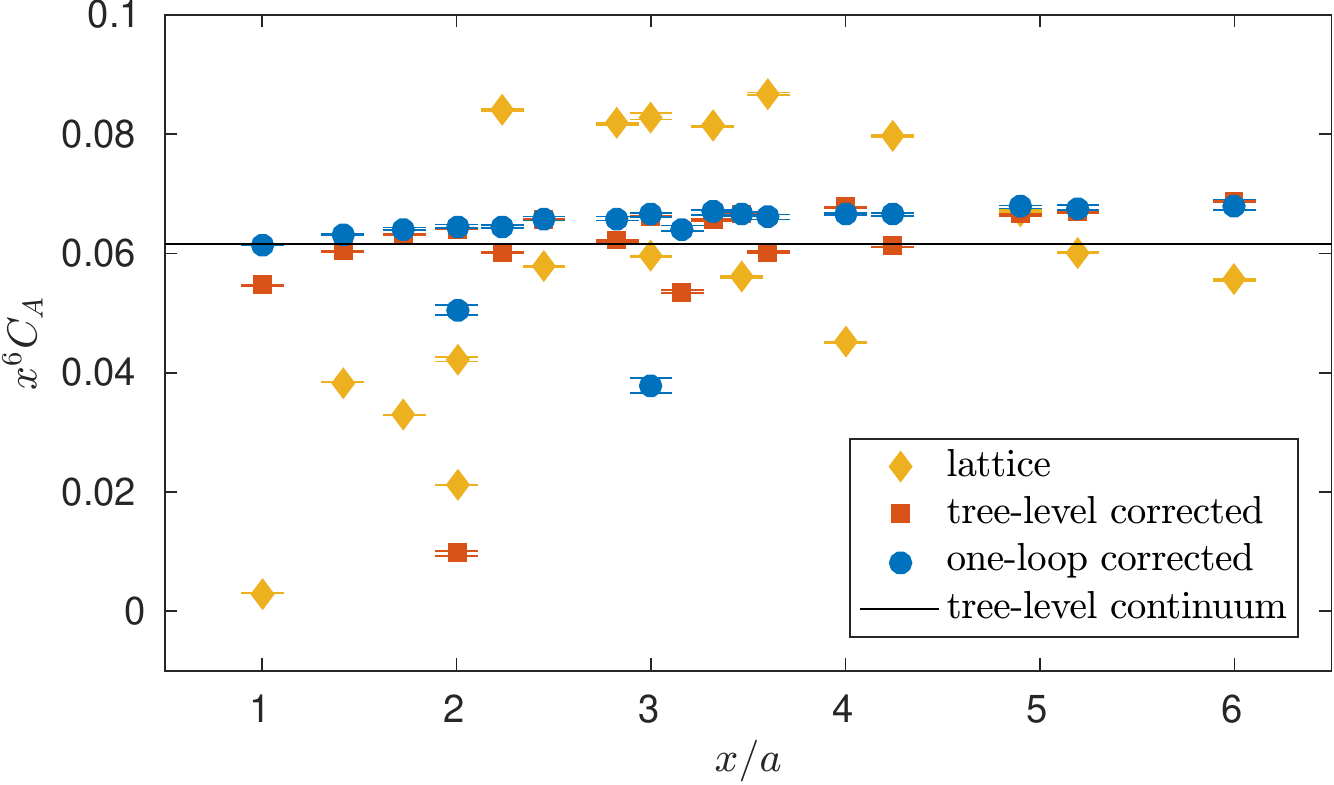}
\caption{Impact of the tree-level (red squares) and one-loop (blue circles) improvement of the massless axial current-current correlation function at $\beta=3.85$. The unimproved lattice correlators are shown as yellow rhombi. \label{fig:1-loop}}
\end{figure}

\subsection{Numerical effects of the tree-level and 1-loop corrections}
We now move on to discussing the numerical effects of subtracting the free-theory and one-loop lattice artefacts.
We start by showing in Fig.~\ref{fig:1-loop} the corrections in the axial vector channel, at our finest lattice spacing corresponding to $\beta=3.85$.
The plot shows the lattice distance dependence of the correlation function multiplied by $x^6$.
The difference of this observable with respect to its tree-level value, $6/\pi^4$, is the measure of $\alpha_s$ in absence of discretization effects.
The yellow rhombi represent uncorrected correlators.
These are subject to huge hypercubic artefacts and it is clear that the latter are much larger than the sought after continuum effects of non-zero $\alpha_s$.
The data become significantly better-behaved after application of the tree-level correction (red squares).
Nevertheless, the spread of points is still much too large to extract $\alpha_s$.
Finally, after applying the 1-loop NSPT correction, we observe further significant reduction of the spread of points (blue circles).
However, as we argued above, for points that break the rotational symmetry most severely the 1-loop correction cannot account for the induced hypercubic artefacts.
This is particularly well seen for the most ``non-democratic'' points in Fig.~\ref{fig:1-loop}, e.g.\ $(0,0,0,2)$, the clear outlier at $x/a=2$, with the maximum possible values of $X^{[4]}$ (1) and the angle $\theta$ (60 degrees).
The influence of the correction for this point should be contrasted with the one for the other point at $x/a=2$, i.e.\ $(1,1,1,1)$, which, in turn, has the smallest possible values of $X^{[4]}$ (0.25) and $\theta$ (0).
In this case, a perfect collapse onto a universal curve formed by several other points is observed.

Thus, we restrict the points used in the analysis to ones that are sufficiently ``democratic'' to allow for a reliable one-loop correction.
In practice, the points that collapse to a universal curve are ones for which at most one component vanishes.
For the remaining points, with two or three zero components, $\theta\geq$ 45 degrees.
This leaves us still with many types of points from which we can hope to extract $\alpha_s$.
However, since we need to match to perturbation theory at not too small energy scales, i.e.\ at relatively small distances, we are limited to points that allow us to reach such distances.
Hence, we concentrate on 3 types of points:
\begin{itemize}
 \item $(0,k,k,k)$,
 \item $(k,k,k,k)$,
 \item $(0,k,k,2k)$,
\end{itemize}
with $k$ being a positive integer.
These are the only types of points that, given the range of our lattice spacings, allow us to reach distances down to the lower edge of our $\alpha_s$ extraction window (0.13 fm) with at least two lattice spacings, thus making them usable in continuum limit fits of the correlators.

\begin{figure}
\includegraphics[width=0.49\textwidth]{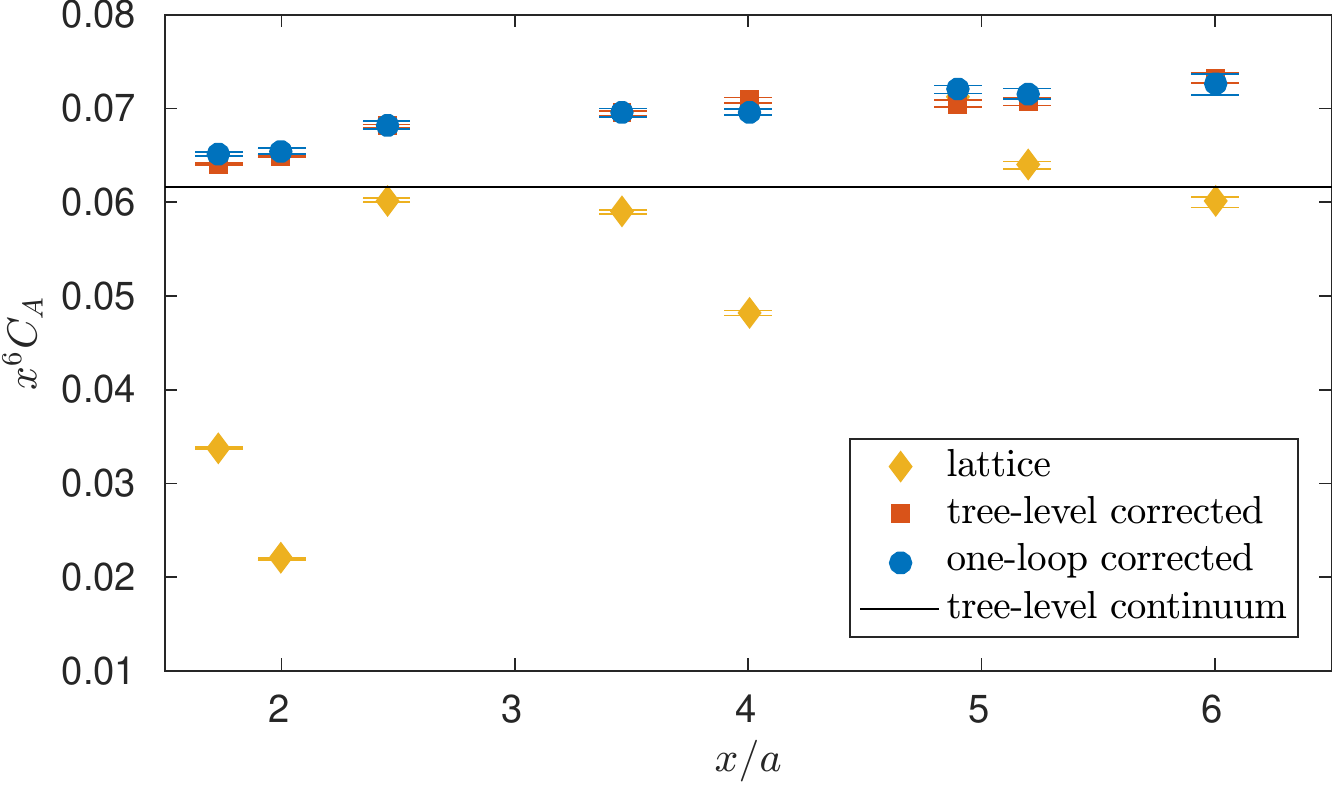}
\includegraphics[width=0.49\textwidth]{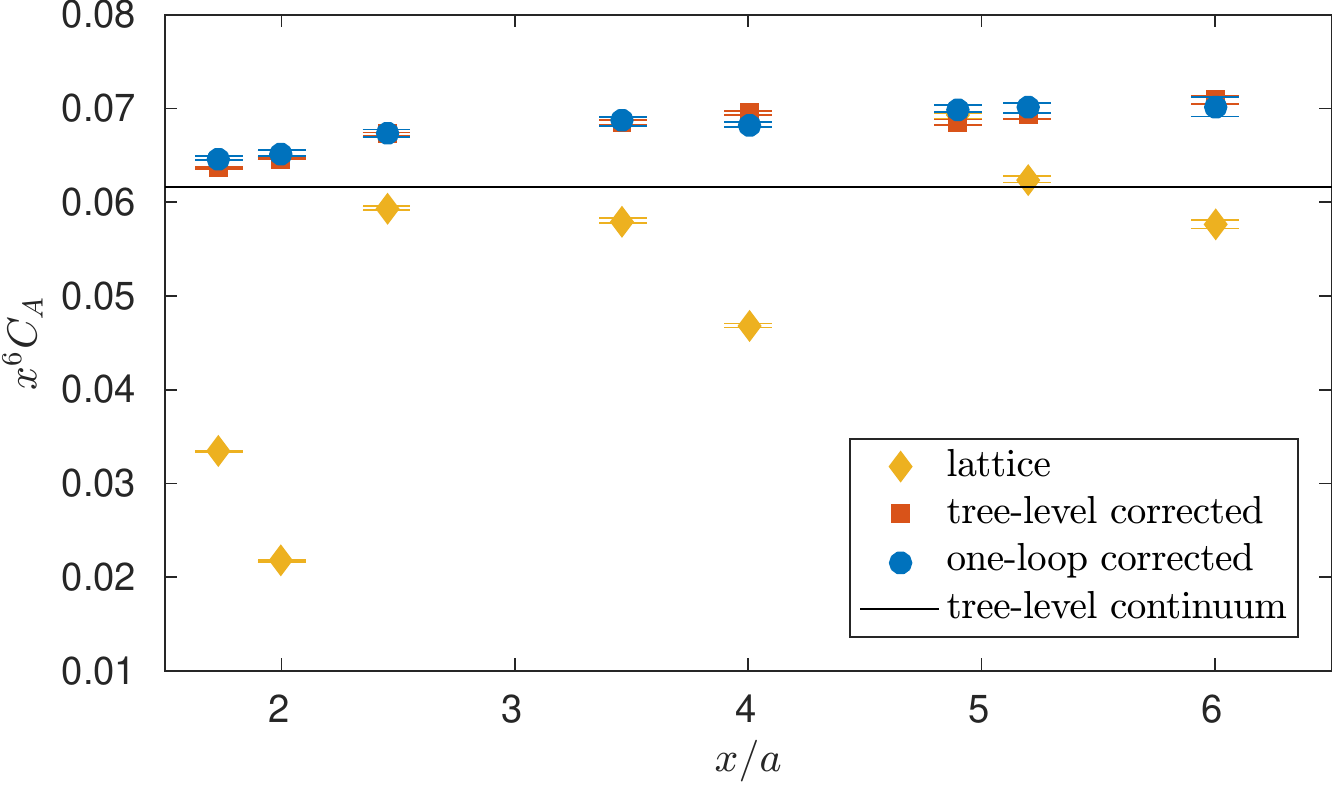}
\includegraphics[width=0.49\textwidth]{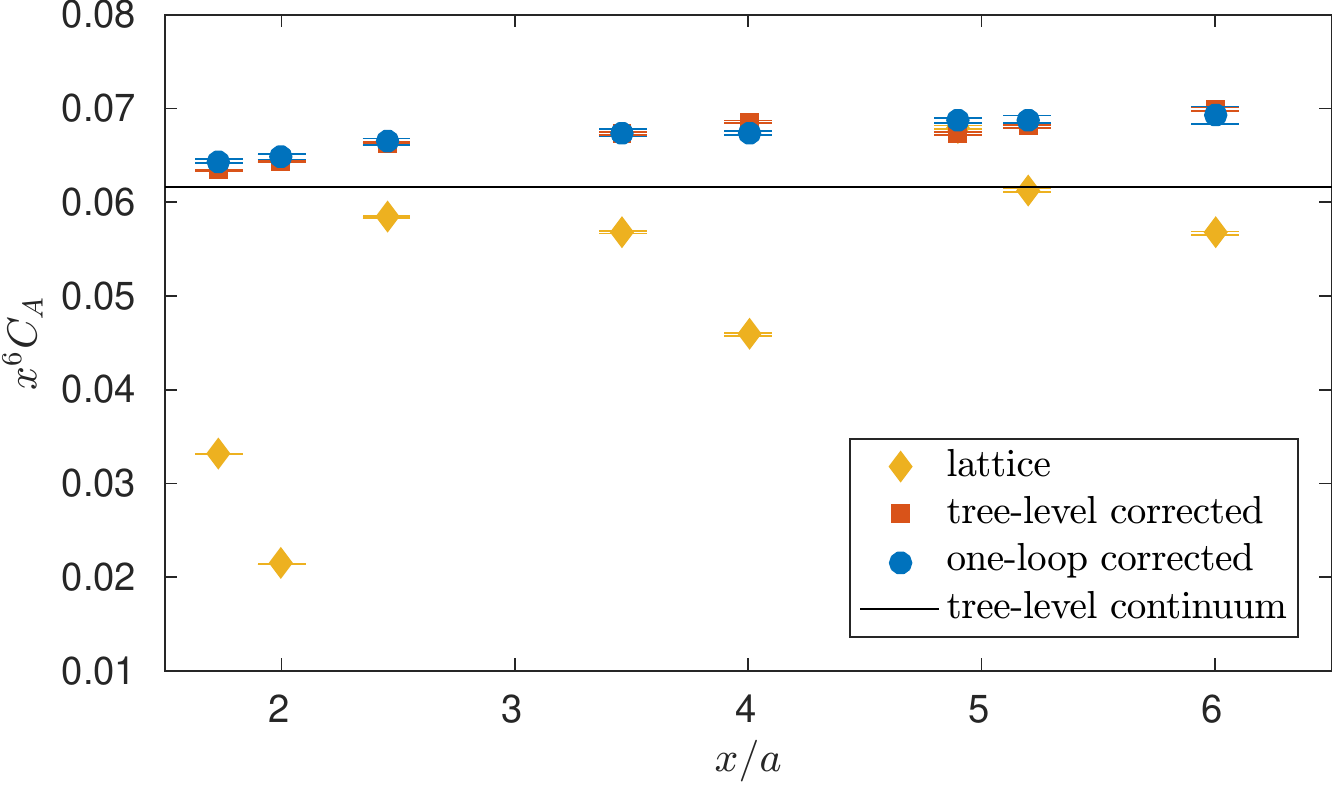}
\includegraphics[width=0.49\textwidth]{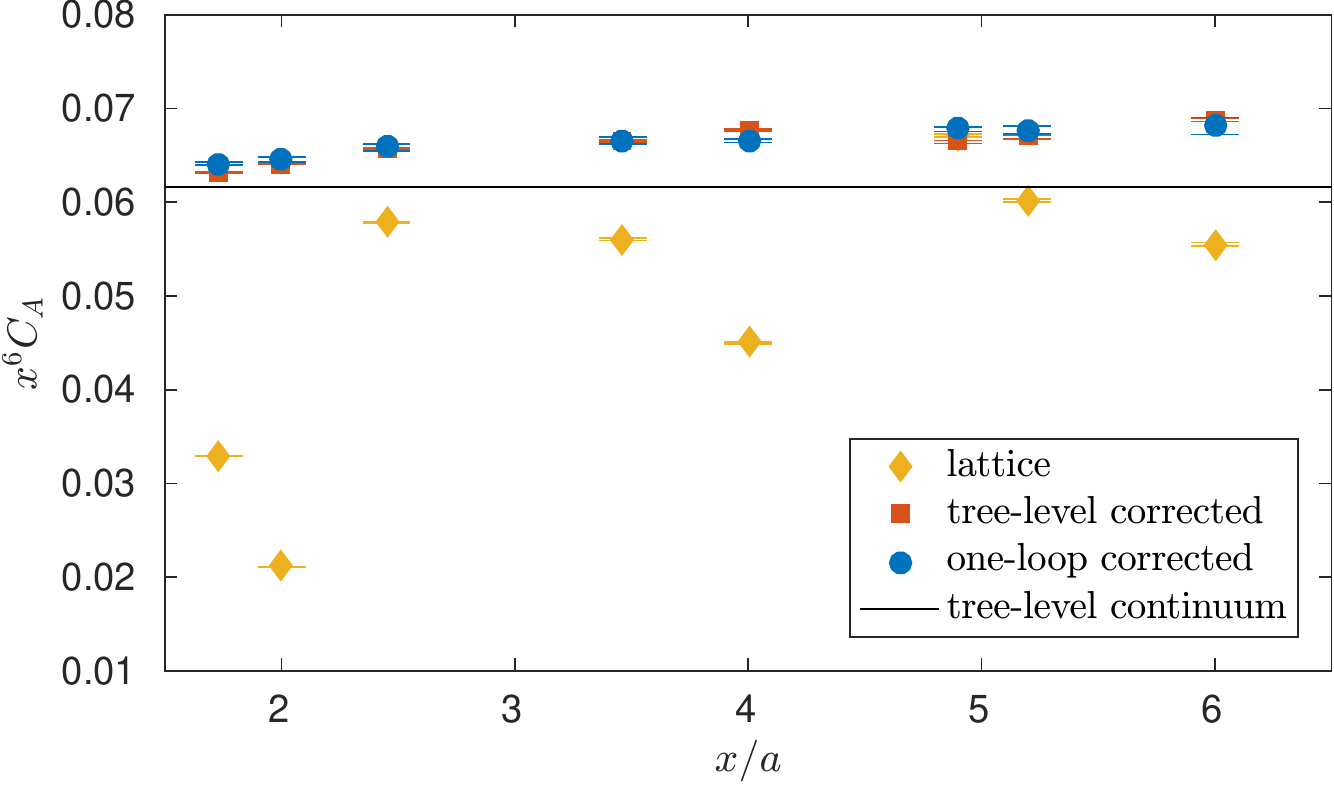}
\caption{Impact of the tree-level (red squares) and one-loop (blue circles) improvement of the massless axial current-current correlation function for types of points used in the extraction of $\alpha_s$. The unimproved lattice correlators are shown as yellow rhombi. Four values of $\beta$: 3.46 (upper left), 3.55 (upper right), 3.70 (lower left) and 3.85 (lower right). \label{fig:1-loop_allbetas}}
\end{figure}

We now show the effect of the tree-level and 1-loop corrections for the 3 types of points used in the extraction of $\alpha_s$, for all 4 lattice spacings that we employ (Fig.~\ref{fig:1-loop_allbetas}).
The unimproved correlators (yellow rhombi) form strikingly similar patterns at all values of $\beta$.
This indicates that the bulk of discretization effects is interaction-independent, i.e.\ scale-independent, and can be considered ``pure'' hypercubic artefacts.
This is confirmed when looking at the effect of the free-theory ($\beta$-independent) correction (red squares).
The distance-dependence of the correlator becomes \emph{almost} smooth.
One may ask at this stage whether the 1-loop subtraction of remaining discretization effects is mandatory in view of this smooth behavior.
The answer to this question will become apparent in the next section of this supplement when we contrast continuum limit fits from tree-level-corrected and from the one-loop-corrected data.
At the scale of Fig.~\ref{fig:1-loop_allbetas}, the data with one-loop perturbative artefacts subtracted (blue circles) do not look very different from their tree-level corrected counterparts.
Nevertheless, the one-loop subtraction is absolutely crucial for a reliable approach to the continuum limit.
This is a similar situation to the one in the above-discussed computations of renormalization functions in momentum space, where such one-loop corrections of hypercubic artefacts have become a standard tool and one that is essential for the overall reliability of this non-perturbative renormalization program.

\section{Continuum limit of the correlation functions}

\begin{figure}
\includegraphics[width=0.40\textwidth]{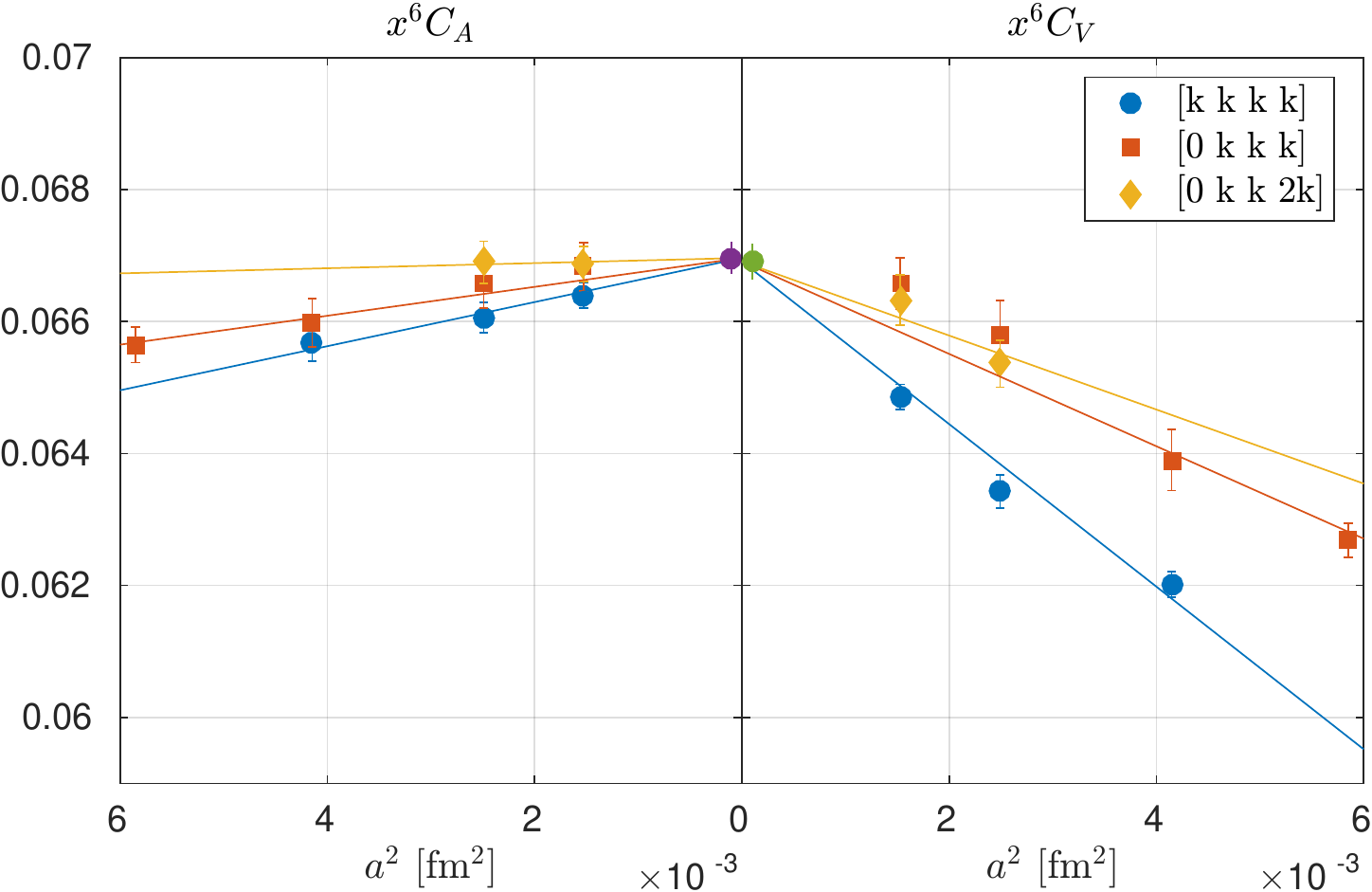}
\includegraphics[width=0.40\textwidth]{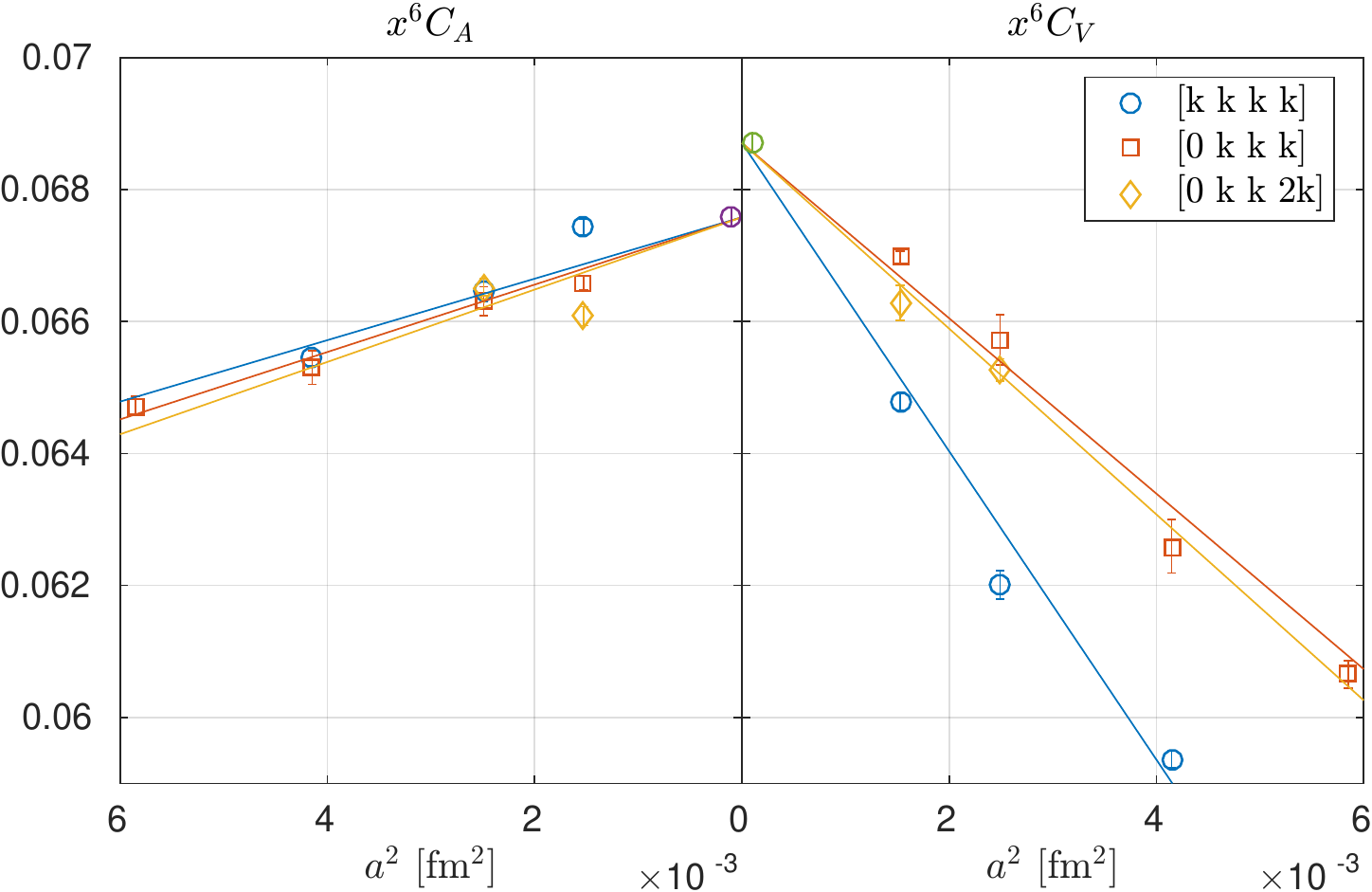}
\includegraphics[width=0.40\textwidth]{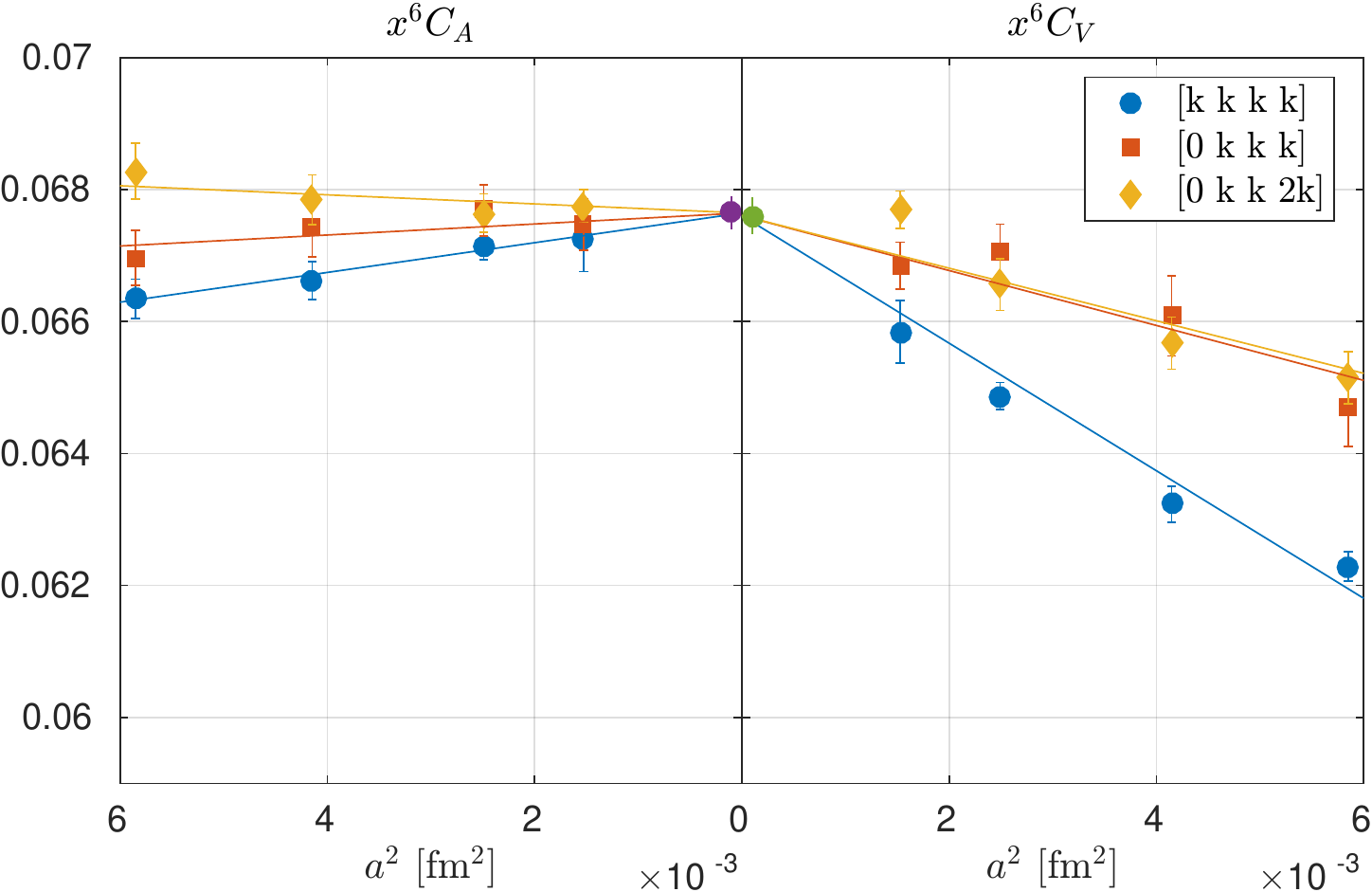}
\includegraphics[width=0.40\textwidth]{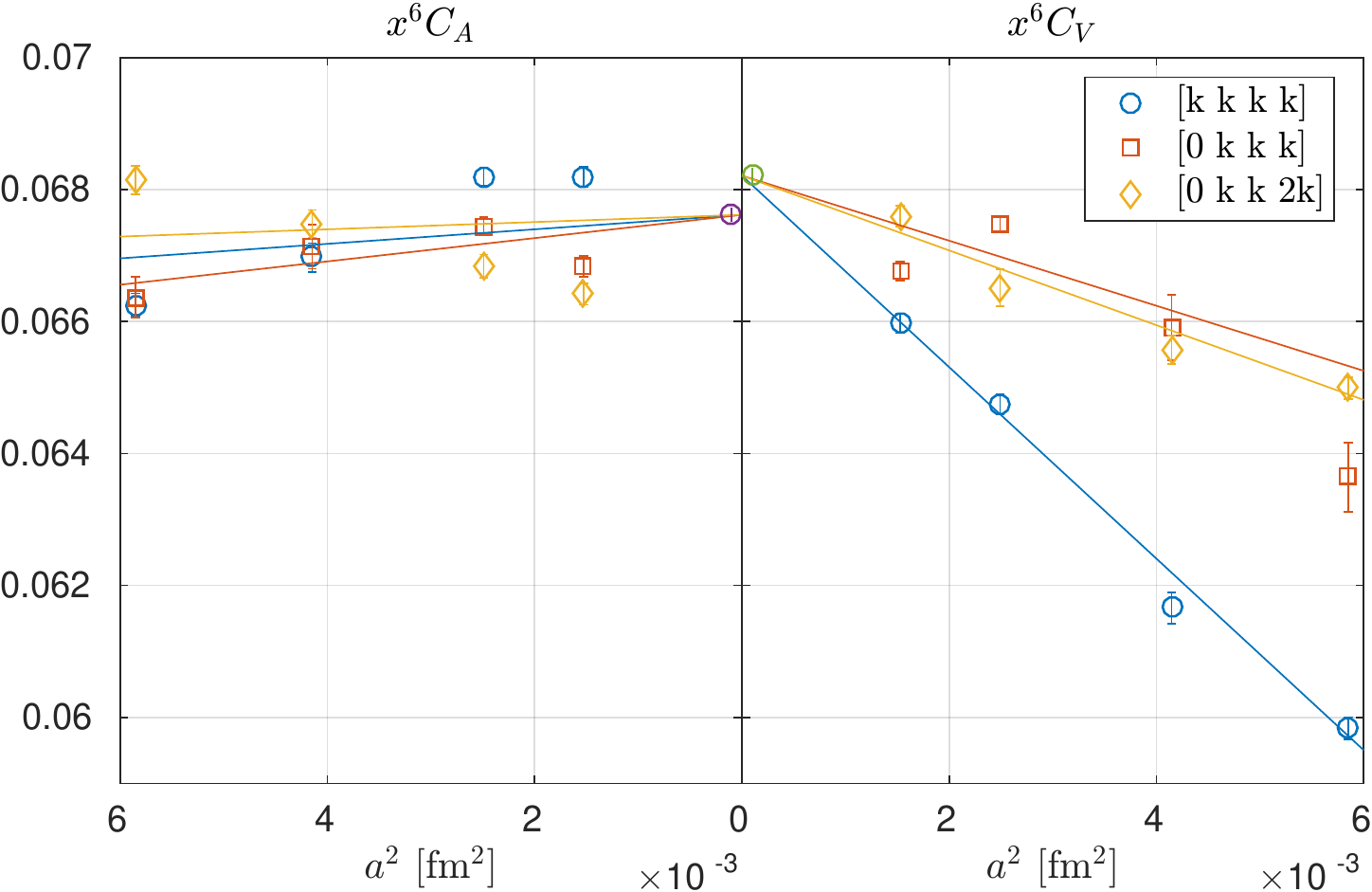}
\includegraphics[width=0.40\textwidth]{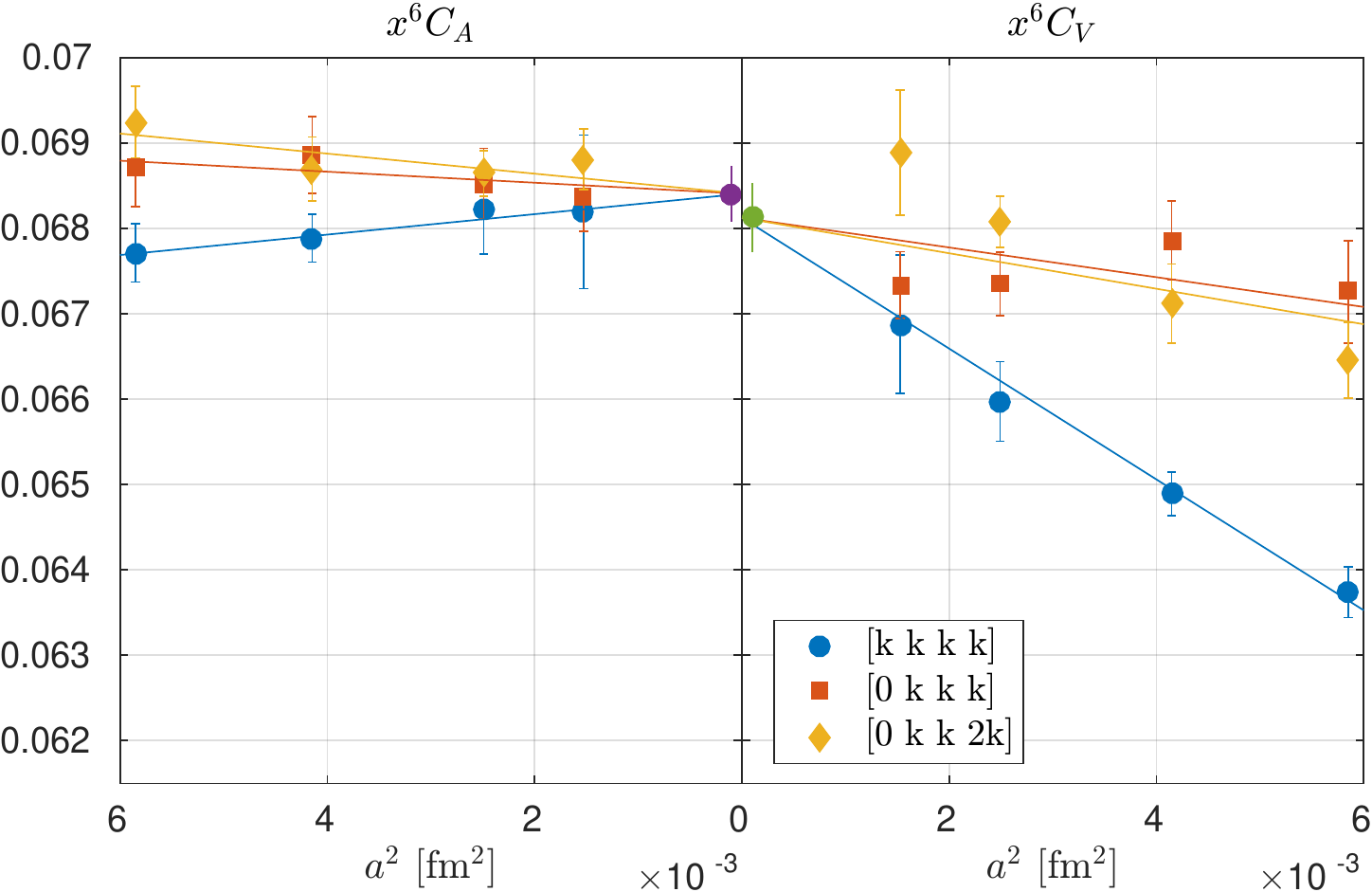}
\includegraphics[width=0.40\textwidth]{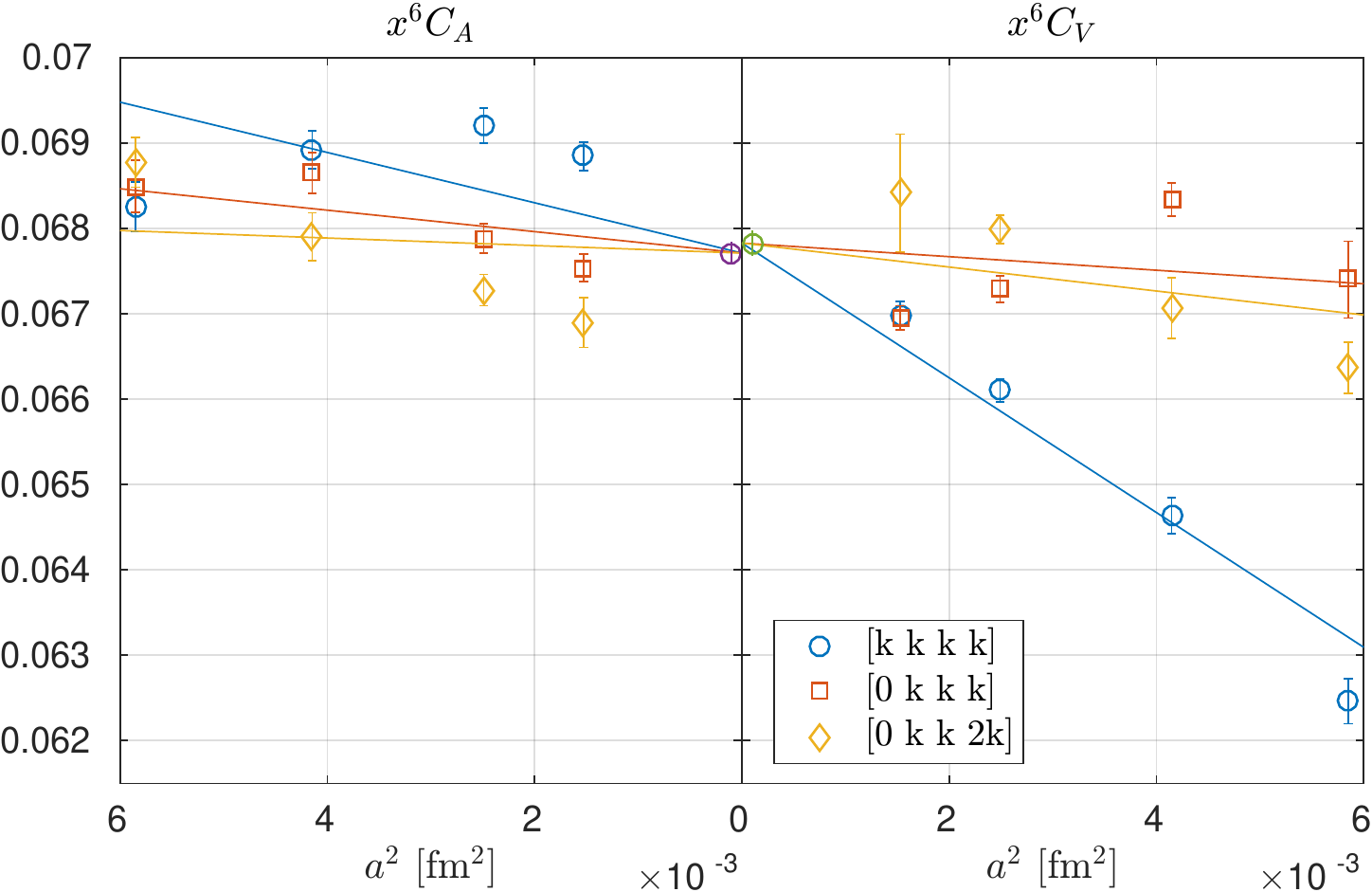}
\includegraphics[width=0.40\textwidth]{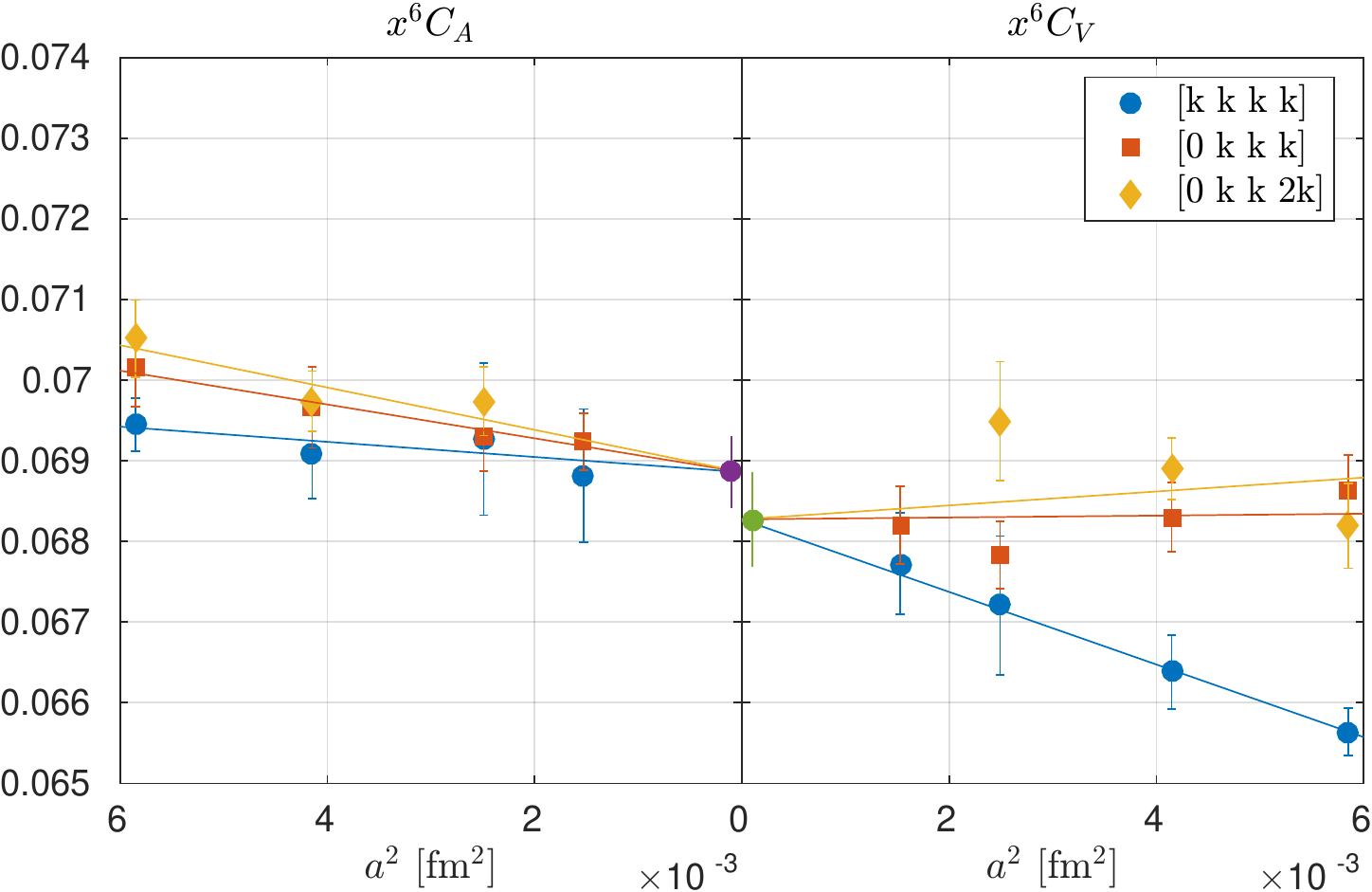}
\includegraphics[width=0.40\textwidth]{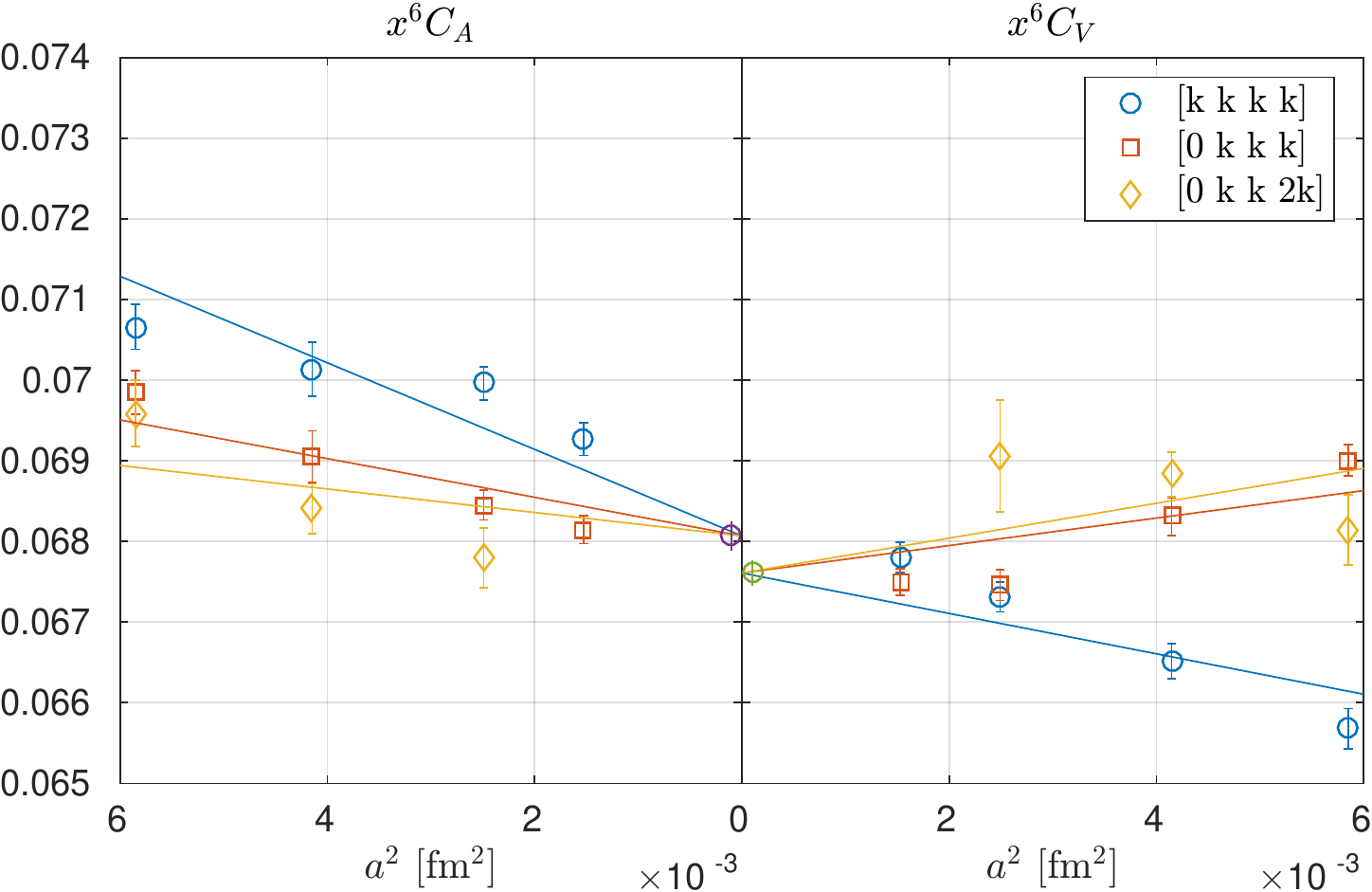}
\includegraphics[width=0.40\textwidth]{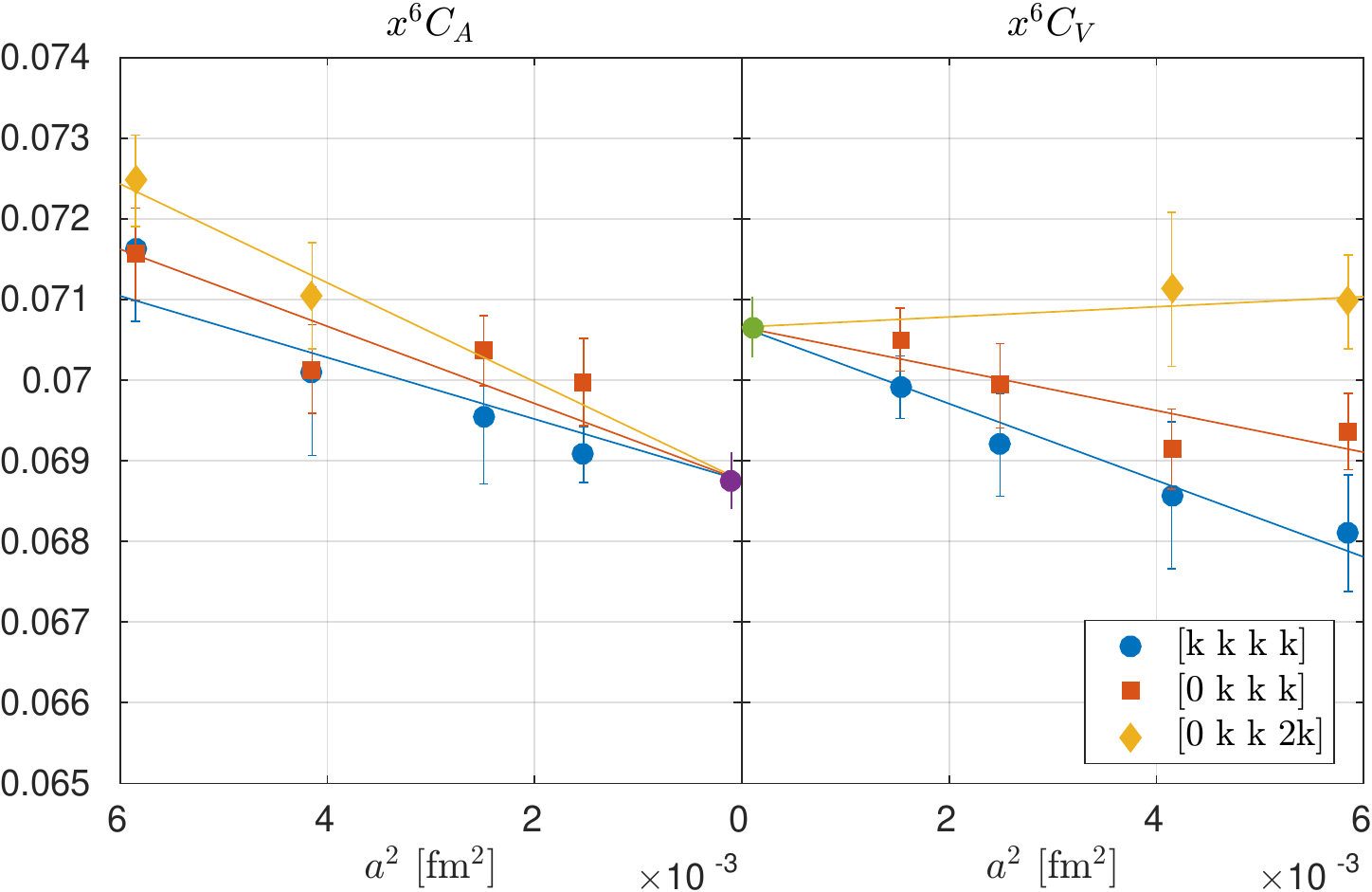}
\includegraphics[width=0.40\textwidth]{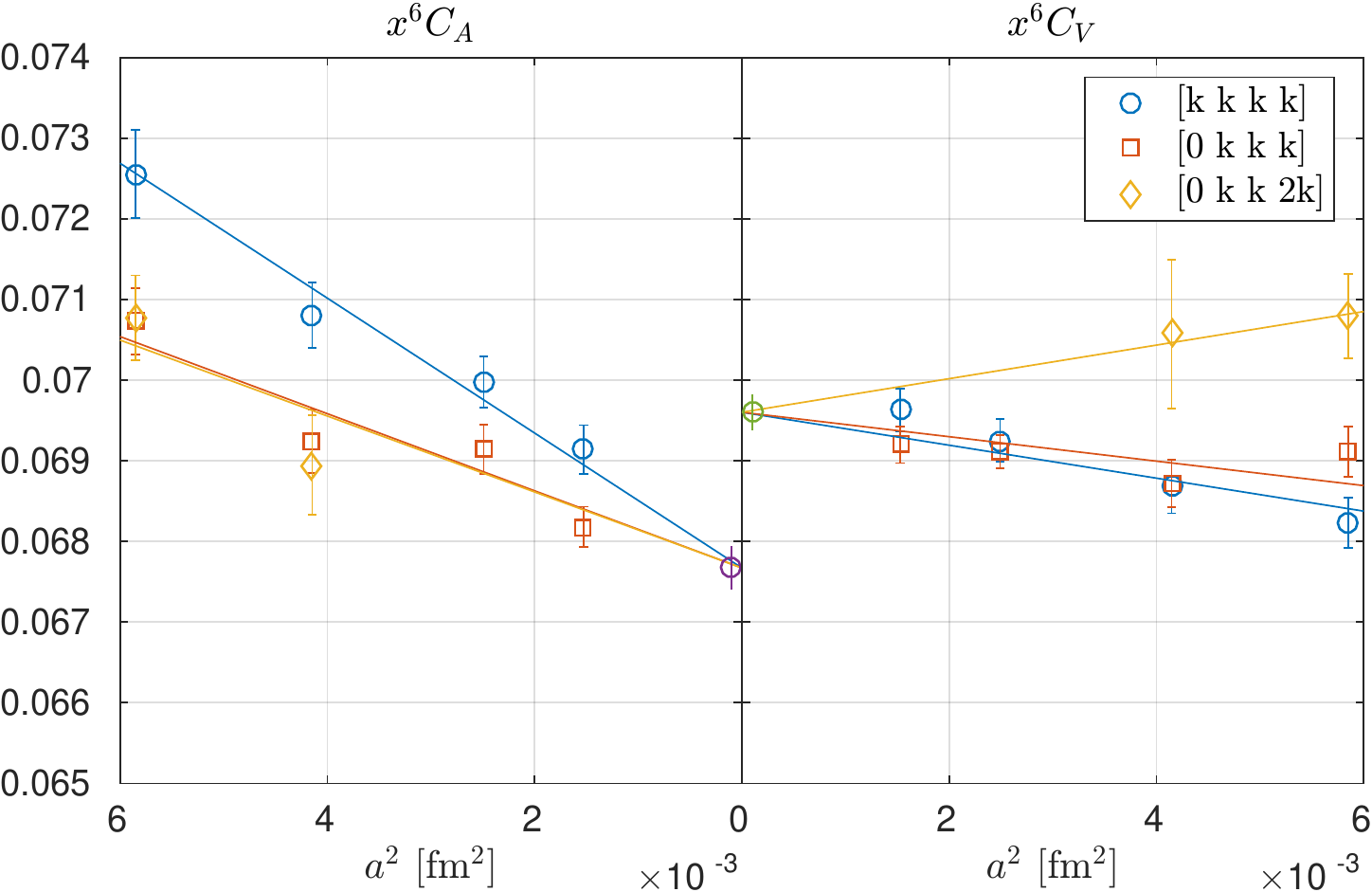}
\caption{Continuum limit extrapolation of the axial vector and vector correlation functions. The left plots show one-loop-corrected correlators and the right plots ones with only the tree-level correction applied. The physical distance is, from to top to bottom: 0.15, 0.19, 0.24, 0.3, 0.4 fm. \label{fig:cont}}
\end{figure}

The crucial issue for a robust extraction of $\alpha_s$ with the coordinate-space method is to obtain reliable continuum extrapolations of the underlying correlation functions.
The data have been extrapolated to the chiral limit, so the only relevant scale left is the correlator distance, $x$.
The lattice-extracted correlators are evaluated at distances corresponding to specific distances, corresponding to multiples of the lattice spacing: $\sqrt{3}ka$ for points of type $(0,k,k,k)$, $2ka$ for $(k,k,k,k)$ and $\sqrt{6}ka$ for $(0,k,k,2k)$, with $k \in \{1,2,3\}$ for the relevant range of distances.
To find the correlator at these distances, we employ the interpolation procedure described in the main part of the paper.
We emphasize that interpolating between the same type of points, which break the continuum rotational symmetry in the same way (reflected in the patterns observed in Fig.~\ref{fig:1-loop_allbetas}), the discretization effects are fixed.

In this way, in each channel, we have evaluations of the correlation function at 4 lattice spacings and corresponding to 3 types of points.
Hence, we have up to 12 points for the axial vector case and, independently, up to 12 points in the vector channel.
It is not possible to reach small distances with all types of points at each lattice spacing and thus, in practice, the number of points is smaller at some distances, e.g.\ it is 9 points for $x\lesssim0.151$ fm 
or 11 for $x\lesssim0.185$ fm.
Lattice correlators from different types of points at a fixed physical distance are just different discretizations of the same continuum correlator and even though they, in general, differ at a finite lattice spacing, they need to agree in the continuum limit.
We checked that this property indeed holds by performing independent continuum extrapolations for each type of point.
After establishing this fact, we performed combined fits to all types of points, independently in the vector and axial vector channels.
In our lattice setup, the correlators are fully $\mathcal{O}(a)$-improved and thus, the leading cutoff effects are at $\mathcal{O}(a^2)$.
According to this, our fitting ansatz for the continuum limit extrapolation takes the form
\begin{equation}
\label{eq:cont1}
C_\Gamma(1/x,a) = C_\Gamma(1/x,a=0) + \sum_{i = \textrm{point type}} \alpha_i a^2, 
\end{equation}
with a common continuum limit $C_\Gamma(1/x,a=0)$ for all types of points and 3 slopes $\alpha_i$ of the $\mathcal{O}(a^2)$ cutoff effects as fitting parameters.
We also consider a fitting ansatz with higher-order ($\mathcal{O}(a^3)$ in our setup) discretization effects included:
\begin{equation}
\label{eq:cont2}
C_\Gamma(1/x,a) = C_\Gamma(1/x,a=0) + \sum_{i = \textrm{point type}} \alpha_i a^2 + \sum_{i = \textrm{point type}} \beta_i a^3, 
\end{equation}
with 3 additional fitting parameters.

In Fig.~\ref{fig:cont}, we show the results of continuum limit extrapolations in both channels for several physical distances: 0.15, 0.19, 0.24, 0.3 and 0.4 fm.
The larger distances are not directly relevant for the extraction of $\alpha_s$, but we are interested in establishing the robustness of the fits in general and in finding the distance where the continuum limits of axial vector and vector correlators are indistinguishable within our precision (see next section of this supplement).
The latter is important to check whether both channels can be ultimately combined to gain in statistical precision.

The left plots of Fig.~\ref{fig:cont} show the fits for the case when the correlation functions are improved with the 1-loop NSPT correction, whereas the right plots show our data for only tree-level-corrected correlators.
Despite the apparently small size of the 1-loop correction as compared to the tree-level correction (see Fig.~\ref{fig:1-loop_allbetas}), the effect of the former is seen to be absolutely crucial at the stage of the continuum fits.
The fitting ansatz (\ref{eq:cont1}) provides a very good description of the NSPT-corrected data, with $\chidof\approx0.1-1.3$ in the axial vector channel and $\approx0.2-2$ in the vector channel, depending on the physical distance.
On the other hand, the fits of Eq.~(\ref{eq:cont1}) for tree-level-corrected correlators fail to provide an acceptable description of data, with $\chidof$ typically in the range $10-20$ for distances smaller than approx.\ 0.25 fm.
Only at distances larger than this value, the tree-level-corrected data start to be well-described by Eq.~(\ref{eq:cont1}), which is particularly well seen in the bottommost plot of Fig.~\ref{fig:cont}, i.e.\ at $x=0.4$ fm.
This is in perfect accordance with the expectation that discretization effects become smaller at larger distances.

As mentioned above, we attempted also the inclusion of higher-order discretization effects in the fitting ansatz.
However, this implies 3 additional fitting parameters and the number of degrees of freedom left in the fits becomes only 2 to 5, depending on the distance.
Thus, this approach fails to offer any improvement in the continuum limit extrapolations, leading to statistically insignificant coefficients $\beta_i$ in Eq.~(\ref{eq:cont2}).
Nevertheless, the quality of fits of Eq.~(\ref{eq:cont1}) is fully satisfactory for the NSPT-corrected data.

\section{Difference between axial vector and vector correlators}
The axial vector and vector correlation functions are related to each other by a flavor non-singlet $SU(2)_A$ chiral symmetry.
This symmetry is broken spontaneously at some energy scale, $\Lambda_\chi$, of the order of a few hundred MeV.
In coordinate space, correlation functions at all distances receive contributions from all energy scales.
Thus, the chiral symmetry is broken for any distance and the axial vector correlator is never equal to the vector correlator.
However, at small distances, the contributions from low energy scales are strongly suppressed and approximate equality of the correlator holds to a good extent.
The difference between the vector and axial vector correlators was estimated in the framework of the operator product expansion (OPE) by Shifman et al.\ \cite{SHIFMAN1979385} many years ago.
They found that the leading-order difference in the chiral limit is proportional to the chiral condensate (squared) and emerges $\propto x^6\ln(\mu_\chi^2 x^2)$, where the constant $\mu_\chi$ needs to be plugged in from experimental data or established in a lattice calculation.
When using an estimate of $\mu$ from Ref.~\cite{PhysRevLett.86.3973}, we find the difference between the two types of correlators of approx.\ 0.03\% at $x=0.1$ fm and it exceeds 1\% at distances larger than approx.\ $0.3$ fm. 

In this work, we test this prediction using our lattice data.
This is a physically interesting test in itself, but apart from this, it serves also the practical purpose of averaging the data from the two channels in the regime of distances relevant for the extraction of $\alpha_s$.
To this aim, equivalence between the two correlators needs to be reliably established, i.e.\ the continuum limits of both correlation functions need to be compatible within statistical errors in the range in question. 

We report the continuum limits of both correlation functions in the main part of the paper (inset of Fig.~2).
Here, we discuss a bit more details, based on the continuum extrapolation plots of Fig.~\ref{fig:cont}.
As we argued in the previous section, it is crucial to use 1-loop corrected data.
With only the tree-level correction, the quality of the continuum extrapolations is unacceptable and no conclusions can be drawn about the equivalence of the two correlators.
The NSPT-corrected data allow us to perform meaningful fits to the continuum limit, with $\chidof$ of order 1.
In the range relevant to the extraction of $\alpha_s$ ($x<0.2$ fm), the statistical error of both correlators is $0.2-0.4\%$ in the continuum limit, while the observed differences in the central values do not exceed $0.1-0.2\%$ (see the two upper rows of Fig.~\ref{fig:cont}).
This is consistent with the expectation based on the value of $\mu_\chi$ from Ref.~\cite{PhysRevLett.86.3973} within a considerable safety margin and thus, we can safely average the two correlators in the range of $\alpha_s$ extraction.
Going beyond this range, the central values of the correlators in the continuum limit start to differ, but the statistical errors increase and both types of correlators are still consistent, within an uncertainty reaching approx.\ $1\%$ at distances around 0.3 fm (see the third and fourth row of Fig.~\ref{fig:cont}).
Thus, our result for the difference between the axial vector and vector correlators is still compatible with zero and, within our precision, also with the number based on OPE.
Increasing the distance further, we find incompatible continuum limits for both correlators for the first time at $x=0.35$ fm and we illustrate this situation with a yet larger distance, $x=0.4$ fm (bottommost row of Fig.~\ref{fig:cont}).
At this distance, we find the correlators differ by approx.\ $3\%$, which is yet again consistent with OPE within our error.

To summarize, the axial vector correlator and the vector correlator differ at any non-zero distance.
The precision of our lattice data allows us to observe this difference at distances larger than approx.\ 0.35 fm.
At such distances, the difference in the continuum limits of both types correlators reaches $2-3\%$ according to our results, which is consistent with predictions based on OPE, with an independent estimate of the value of the input parameter $\mu_\chi$.
Our findings are also qualitatively and quantitatively consistent with the lattice study of Ref.~\cite{Tomii:2017cbt} (see Fig.~13 of this paper) and with the experimental data on hadronic $\tau$ decays from the ALEPH Collaboration \cite{Davier:2013sfa}.
We note that the distances where the two correlators are found to differ are not related in any way to the breakdown of perturbation theory, but rather indicate where the effects of spontaneously broken chiral symmetry become sizable in coordinate space observables.

\twocolumngrid
\bibliography{references2}
\end{document}